\documentclass[journal]{IEEEtran}

\usepackage{cite}
\usepackage{amsmath,amssymb,amsfonts}
\usepackage{textgreek}
\usepackage{algorithmic}
\usepackage{graphicx}
\usepackage[caption=false,font=normalsize,labelfont=sf,textfont=sf]{subfig}
\usepackage{textcomp}
\usepackage{flushend}
\usepackage{xcolor}
\def\BibTeX{{\rm B\kern-.05em{\sc i\kern-.025em b}\kern-.08em
    T\kern-.1667em\lower.7ex\hbox{E}\kern-.125emX}}

\usepackage{tipa}
\usepackage[linesnumbered,ruled,vlined]{algorithm2e}

\begin{document}
\pagenumbering{gobble}
\title{\huge{Data-Aided Bayesian Learning for CSI Estimation over Doubly-Selective DCO-OTFS MIMO VLC Channels with Affine-Precoded Superimposed Training Sequences}}
\author{
Shubham~Saxena,~\IEEEmembership{Graduate Student Member,~IEEE,} Suraj~Srivastava,~\IEEEmembership{Member,~IEEE,} Aditya~K.~Jagannatham,~\IEEEmembership{Senior Member,~IEEE,} and Lajos~Hanzo, ~\IEEEmembership{Life Fellow,~IEEE}\vspace{-8mm} 
\thanks{Shubham Saxena, and Aditya K. Jagannatham are with the Department of Electrical Engineering, Indian Institute of Technology Kanpur, Kanpur-$208016$, India (e-mail: \{shubs20;  adityaj\}@iitk.ac.in). Suraj Srivastava is with the Department of Electrical Engineering, Indian Institute of Technology Jodhpur, Rajasthan $342030$, India (email: surajsri@iitj.ac.in). L. Hanzo is with the School of Electronics and Computer Science, University of
Southampton, Southampton SO$17$ $1$BJ, U.K. (email: lh@ecs.soton.ac.uk).
}
}

\maketitle

\vspace{-10mm}
\begin{abstract}
An orthogonal affine-precoded superimposed training sequence (AP-STS)-based framework is conceived for cyclic-prefix (CP)-assisted multiple-input multiple-output (MIMO) direct-current-biased orthogonal time frequency space (DCO-OTFS) visible light communication (VLC) links using arbitrary transmit-receive pulse shaping for transmission over doubly selective channels. For each light-emitting diode (LED), the pilot and data matrices are jointly affine-precoded and overlaid in the delay-Doppler (DD) domain. Then, a unified end-to-end DD-domain input-output relationship is derived. At each photodiode (PD), orthogonal precoders are utilized to separate the pilot and data components, thereby suppressing mutual interference. Building on this model, an expectation-maximization (EM) driven DD-domain pilot-aided Bayesian learning (DD-PBL) scheme is developed to estimate the channel state information (CSI). A DD-domain data-aided Bayesian learning (DD-DBL) procedure is then proposed for jointly refining the CSI and detecting data by exploiting the detected symbols as virtual pilots in the spirit of decision-directed channel estimation. The linear minimum mean square error (LMMSE) detector harnessed explicitly accounts for CSI uncertainty due to realistic estimation errors. In addition, Bayesian Cramér-Rao lower bounds (BCRLBs) are derived for the MIMO DCO-OTFS VLC setting considered. Numerical results confirm improved normalized mean-square-error (NMSE), reduced pilot overhead, and mitigated symbol error-rate (SER) relative to recent benchmarks.

\end{abstract}

\begin{IEEEkeywords}
Affine precoded, Bayesian learning (BL), BCRLB, delay-Doppler domain channel, optical OTFS, superimposed training sequences, visible light communication.
\end{IEEEkeywords}
\vspace{-3mm}

\section{Introduction}

\IEEEPARstart{C}{onventional} wireless technologies, including radio frequency (RF) and microwave-based links, have historically supported ubiquitous connectivity in mobile access networks and related wireless systems. Nevertheless, the rapid proliferation of wireless devices and the persistent demand for higher data rates have intensified concerns regarding RF spectrum congestion. In this setting, visible light communication (VLC), as a form of optical wireless communication (OWC), has emerged as a complementary wireless paradigm that leverages light-emitting diode (LED)-based transmitters and photodiode receivers for data transfer \cite{11359272,11417983,rahman2023channel}. VLC conveys information by modulating the optical intensity, thereby exploiting the wide visible light band ($430$ to $790$ terahertz) as a largely underutilized resource for high-capacity short-range links. By mapping binary information to variations in light intensity, VLC can provide high throughput with low electromagnetic interference, improved security, and favorable energy efficiency \cite{xu2023optical,rahman2023channel,saxena2025multiple,mushfique2020optimization}. Its compatibility with the indoor lighting infrastructure, together with the absence of RF licensing requirements, further enhances its practical appeal, making VLC a promising solution for reliable and secure wireless access.

The VLC propagation medium generally comprises both line-of-sight (LoS) and non-LoS (NLoS) components. The LoS component corresponds to the direct optical path spanning from the LED to the photodetector (PD), whereas NLoS components arise due to reflections from walls, objects, and other surfaces. Multipath VLC propagation has been studied in \cite{11417983,chen2016adaptive,schulze2024dispersive,you2019performance}. The impact of higher-order reflections was further analyzed in \cite{zhou2014impact}, which indicates that channel models accounting for only a limited number of reflections may be insufficient for high-rate VLC links. In a related vein, \cite{saxena2025multiple} considers a multipath VLC model comprising both LoS and NLoS components. Such multipath effects induce temporal dispersion in the received signal, and the corresponding power delay profile is investigated by Lee \textit{et al.} \cite{lee2011indoor}. The resultant delay spread produces inter-symbol interference (ISI) in indoor VLC systems \cite{9223722,barry1993simulation}. To alleviate ISI, optical orthogonal frequency division multiplexing (O-OFDM) has been widely studied as an effective multicarrier modulation technique for LED-based VLC, offering high spectral efficiency while mitigating ISI \cite{azim2019energy,you2019performance}. Moreover, intensity modulation with direct detection (IM/DD) is commonly adopted for O-OFDM reception due to its implementational simplicity \cite{you2019performance,saxena2023sparse}.

Concurrently, orthogonal time frequency space (OTFS) modulation has gained significant attention for wireless links affected by doubly selective channels, particularly in mobility-oriented scenarios \cite{hadani2017orthogonal}. Unlike orthogonal frequency division multiplexing (OFDM), which can suffer from Doppler-induced intercarrier interference (ICI), OTFS represents symbols on a delay-Doppler (DD) grid, improving robustness to time-frequency (TF) selectivity and supporting more reliable channel estimation (CE). Although DD-domain processing introduces additional complexity, OTFS has demonstrated consistent advantages even under static multipath conditions \cite{raviteja2019otfs}. These observations motivate the employment of optical OTFS (O-OTFS) for VLC \cite{zhong2020orthogonal,zheng2021dco}, especially in high-rate optical wireless links where multipath dispersion and time-frequency selectivity may limit the effectiveness of conventional waveform design. In such systems, the achievable rate is strongly influenced by the accuracy of channel state information (CSI). Since conventional CE procedures may incur high pilot overhead and can be less effective in doubly selective optical environments, recent research has focused on more efficient CSI acquisition strategies. The most pertinent contributions are reviewed next.
\vspace{-3mm}
\subsection{Literature Review}
Since VLC employs IM/DD, the transmitted waveform must be real-valued and non-negative, which necessitates careful optical adaptations of OTFS. Zhong \textit{et al.} \cite{zhong2020orthogonal} developed direct current-biased optical OTFS (DCO-OTFS) by enforcing two-dimensional Hermitian symmetry and demonstrated improved bit error-rate (BER) together with reduced peak-to-average power ratio (PAPR), when compared to direct current-biased O-OFDM (DCO-OFDM). Zheng \textit{et al.} \cite{zheng2021dco} incorporated DCO-OTFS into a full-duplex relay-assisted VLC architecture and reported that the cyclic-prefix (CP) overhead can be reduced relative to DCO-OFDM, leading to improved spectral efficiency. To address front-end non-idealities, Sharma \textit{et al.} \cite{sharma2023hyperparameter} proposed a hyperparameter-free receiver based on random Fourier features to mitigate the LED's nonlinearity in O-OTFS VLC links. In \cite{sinha2021otfs}, the authors conceived O-OTFS modulation for static indoor dual-LED VLC scenarios. In a related study \cite{sinha2021quad}, O-OTFS was further investigated in static indoor quad-LED VLC configurations. Both \cite{sinha2021otfs,sinha2021quad} reported that O-OTFS achieves superior performance relative to O-OFDM. Furthermore, Xu \textit{et al.} \cite{xu2023optical} presented an O-OTFS framework emphasizing the bandwidth, power, and energy efficiency benefits over O-OFDM relying on DD-domain CSI estimation. In a related direction, Wang \textit{et al.} \cite{wang2025spectrally} derived a general DD to time-domain (TD) conversion expression using the discrete Zak transform (DZT) and then formulated a maximum likelihood (ML) detection criterion in the DD domain. More recently, Cai \textit{et al.} \cite{cai2025power} proposed a power- and spectrum-efficient O-OTFS design for underwater VLC, whereas Chen \textit{et al.} \cite{chen2025optical} investigated an IM/DD satellite optical link using DCO-OTFS to combat fading and Doppler effects.
\begin{table*}[t]
\centering
\caption{Boldly contrasting our contributions to the literature}
\label{SoTA}
\resizebox{\linewidth}{!}{%
\begin{tabular}{|l|c|c|c|c|c|c|c|c|c|c|c|c|c|c|c|}
\hline
\textbf{Features} & \cite{rahman2023channel} & \cite{xu2023optical} & \cite{saxena2023sparse} & \cite{zhong2020orthogonal} & \cite{zheng2021dco} & \cite{sharma2023hyperparameter} & \cite{sinha2021quad} & \cite{wang2025spectrally} & \cite{cai2025power} & \cite{chen2025optical} & \cite{liao2023sparse} & \cite{mishra2021iterative} & \cite{yuan2021data} & \cite{estrada2019superimposed} & \textbf{Proposed} \\
\hline
O-OTFS & & \checkmark &  & \checkmark & \checkmark & \checkmark & \checkmark & \checkmark & \checkmark & \checkmark & \checkmark &  &  &  & \checkmark \\ \hline
STS & & &  &  &  &  &  &  &  &  &  & \checkmark & \checkmark & \checkmark & \checkmark \\ \hline
MIMO VLC & \checkmark& &  &  &  &  & \checkmark &  &  &  &  &  &  & \checkmark & \checkmark \\ \hline
Data-aided CE & & &  &  &  &  &  &  &  &  &  & \checkmark & \checkmark &  & \checkmark \\ \hline
BL-based CE & & & \checkmark &  &  &  &  &  &  &  & \checkmark &  &  &  & \checkmark \\ \hline
BCRLB &  &  & \checkmark &  &  &  &  &  &  &  &  &  &  &  & \checkmark \\ \hline
Affine-Precoding & & &  &  &  &  &  &  &  &  &  &  &  &  & \checkmark \\ \hline
Fractional Doppler & & &  &  &  &  &  &  &  &  &  &  &  &  &  \checkmark \\ \hline
CSI uncertainty for data detection & & &  &  &  &  &  &  &  &  &  &  &  &  & \checkmark \\ \hline
LMMSE-based modified detection rule & & &  &  &  &  &  &  &  &  &  &  &  &  & \checkmark \\ \hline
\end{tabular}%
}\vspace{-3mm}
\end{table*}

Reliable CSI acquisition is particularly important in VLC, because LoS and NLoS components jointly induce delay dispersion, while mobility can introduce Doppler spread, resulting in doubly selective channels that challenge both estimation and detection. For optical OFDM, Saxena \textit{et al.} \cite{saxena2023sparse} developed Bayesian learning (BL) based CSI estimation for frequency-selective DCO-OFDM and asymmetrically
clipped optical OFDM (ACO-OFDM)-based VLC models, demonstrating notable gains relative to classical estimators. In the OTFS-VLC setting, Liao \textit{et al.} \cite{liao2023sparse} established a DD-domain input-output relationship between pilot observations and the channel, and proposed a BL estimator relying on embedded pilots. Although embedded-pilot structures alleviate the need for a dedicated pilot frame, they still allocate substantial resources to pilots and protective regions, which motivates the employment of more bandwidth-efficient training mechanisms, especially for multiple-input multiple-output (MIMO) VLC schemes.

The RF OTFS literature provides complementary insights into DD-domain modeling and pilot overhead tradeoffs. Impulse-based pilot strategies, including their MIMO extensions, separate pilots across transmit antennas to obtain interference-free DD observations. Yet, they typically require large guard regions and may rely on empirically tuned thresholds, both of which can inflate the overheads and compromise robustness \cite{ramachandran2018mimo}. Embedded pilot and data patterns reduce overhead by sharing a single OTFS frame, but the combined pilot and guard structures can still incur throughput erosion \cite{hadani2017orthogonal}.

To further improve training efficiency, superimposed training sequence (STS) aided techniques embed pilots directly into data \cite{mishra2021iterative,yuan2021data,muntane2024optimal}. In VLC, Estrada \textit{et al.} \cite{estrada2019superimposed} investigated least squares (LS) based CSI estimation for a multiple-input single-output (MISO) DCO-OFDM system using direct pilot-data superposition. Nevertheless, direct superposition inherently creates mutual pilot-data interference, and its mitigation frequently depends on maximum \textit{a posteriori} processing or message passing-based detection, while requiring prior information such as channel quality order or sparsity level, which can be impractical \cite{mishra2021iterative,yuan2021data}. Affine-precoded superimposed training sequences (AP-STS) provide a structured design alternative by applying orthogonal affine precoders so that the pilot and data components can be separated at the receiver through algebraic projection, perfectly eliminating mutual interference across all signal-to-noise ratios. Prior research has studied affine precoding in RF settings \cite{tran2008orthogonal}; however, most existing formulations focus on time-invariant channels and do not directly address doubly selective operation. Furthermore, they remain oblivious to the IM/DD constraints and optical front-end characteristics intrinsic to VLC. As a remedy, data-aided processing can strengthen AP-STS by treating the detected data symbols as reliability-weighted virtual pilots and iteratively refining CSI within a BL framework.

Motivated by the above knowledge gaps, this work develops a data-aided BL framework for CSI estimation in doubly selective DCO-OTFS MIMO VLC systems employing AP-STS signaling. The proposed methodology leverages DD-domain sparsity associated with a limited number of dominant reflectors, improving estimation accuracy without requiring prior statistical knowledge of the channel quality order or sparsity level. By combining interference-free pilot extraction facilitated by orthogonal affine precoding using probabilistically weighted data-aided updates, the proposed method achieves both reduced normalized mean-square-error (NMSE) as well as mitigated symbol error-rate (SER), despite requiring reduced pilot overhead. In addition, the proposed framework is benchmarked against classical sparse recovery baselines, including orthogonal matching pursuit (OMP) and FOCal Underdetermined System Solver (FOCUSS) \cite{saxena2023sparse}. It is evaluated relative to the Bayesian Cramer-Rao lower bound (BCRLB). Table \ref{SoTA} boldly contrasts this work to the representative prior studies, while the following subsection outlines the main contributions in detail.
\vspace{-3mm}
\subsection{Contributions}
\begin{enumerate}
\item An AP-STS signaling model is established for CP-aided MIMO DCO-OTFS using arbitrary transmit-receive pulse shaping. Due to the orthogonality of the affine precoders, the pilot and data components can be disentangled at the receiver through post-multiplication with the associated precoder matrices, which effectively mitigates pilot-data interference.

\item A DD-domain pilot-assisted BL method (DD-PBL) is introduced for MIMO DCO-OTFS that exploits DD-domain sparsity to improve CSI estimation fidelity. The proposed expectation-maximization (EM)-driven iterative hyperparameter learning avoids manual parameter tuning and exhibits stable convergence, which in turn supports improved spectral efficiency.

\item Additionally, a data-aided joint CSI estimation and data detection scheme, denoted as DD-DBL, is developed. By coupling a modified data decision rule with the BL framework, the method performs joint CE as well as data detection and achieves more accurate CSI than pilot-only based approaches. Subsequently, a linear minimum mean square error (LMMSE)-based detection rule is formulated that incorporates both the estimated CSI and the covariance of the CE error, thereby improving detection robustness and enhancing reliability.

\item Closed-form BCRLB results are derived for the analytical performance characterization of the proposed DD-PBL and DD-DBL estimators. The advantages are further verified across a range of operating conditions using NMSE, pilot length, and SER as the primary evaluation criteria.

\end{enumerate}
\vspace{-5mm}
\subsection{Organization}
The remainder of this paper is organized as follows. Section II introduces the data-aided AP-STS MIMO DCO-OTFS VLC system model. Section III develops the sparse DD-domain CIR estimation formulation for the AP-STS MIMO DCO-OTFS VLC framework considered. Section IV details the proposed CSI estimation scheme, DD-PBL, and subsequently presents the data-aided joint CSI estimation and data detection method, DD-DBL, together with the corresponding BCRLB derivation for the MIMO DCO-OTFS VLC system. Section VI reports our simulation results, and Section VII provides our concluding remarks.

\textit{Notations:} 
Notational conventions are as follows: $\mathrm{blkmtx}(\mathbf{A}_1, \mathbf{A}_2, \ldots, \mathbf{A}_N)$ refers to a block-diagonal matrix whose principal diagonal is populated by the matrices $\mathbf{A}_1, \mathbf{A}_2, \ldots, \mathbf{A}_N$, each of which may 
be rectangular. The superscripts $(\cdot)^{T}$, $(\cdot)^{H}$, $(\cdot)^{*}$, and $(\cdot)^{-1}$, are employed to indicate the transpose, conjugate transpose, complex conjugate, and matrix inverse operations, respectively. The symbols $\otimes$ and $\mathrm{Tr}(\cdot)$ designate the Kronecker product and the matrix trace operator, respectively, while $||\cdot||_2$ and $||\cdot||_{F}$ stand for the Euclidean norm and the Frobenius norm, respectively. The operator 
$\mathbb{E}\{\cdot\}$ signifies statistical expectation. Throughout this work, boldface lowercase and uppercase symbols represent column vectors and matrices, respectively. The vectorization operator $\mathrm{vec}(\mathbf{A})$ forms a 
column vector by sequentially concatenating the columns of $\mathbf{A}$, and its inverse $\mathrm{vec}^{-1}(\mathbf{a})$ reconstructs the corresponding matrix. Furthermore, the identity $\mathrm{vec}(\mathbf{ABC}) = 
\left(\mathbf{C}^{T} \otimes \mathbf{A}\right)\mathrm{vec}(\mathbf{B})$ is utilized throughout, where $\otimes$ denotes the Kronecker product.

\vspace{-3mm}
\section{Data-Aided AP-STS MIMO DCO-OTFS VLC System Model}
Consider an AP-STS MIMO DCO-OTFS VLC system characterized by the frame duration $T_d = NT$ and bandwidth $B = M\Delta f$, where $T$ and $\Delta f$ denote the symbol period and subcarrier spacing, respectively, satisfying $T\Delta f=1$. The parameters $N$ and $M$ specify the number of symbols along the time and frequency dimensions of the TF-grid. The corresponding DD grid is discretized with sampling intervals of $\Delta\nu=\frac{1}{T_d}$ and $\Delta\tau=\frac{1}{B}$. The MIMO system employs $N_t$ transmit LEDs and $N_r$ receive PDs.
\vspace{-3mm}
\subsection{Data-Aided AP-STS MIMO DCO-OTFS Modulation}
For the $t$th LED, $1\le t\le N_t$, let $N$ be even and define the set of independent Doppler indices as $\mathcal{K}_{\mathrm{a}}=\left\{1,2,\ldots,\frac{N}{2}-1\right\}$ and $N_{\mathrm{a}}=\left|\mathcal{K}_{\mathrm{a}}\right|=\frac{N}{2}-1.$
Let the data and pilot symbol matrices be $\mathbf{S}_{t}^{d}\in\mathbb{C}^{M\times K_1}$ and $\mathbf{S}_{t}^{p}\in\mathbb{C}^{M\times K_2}$, respectively, where typically $K_1+K_2=N_{\mathrm{a}}$. The elements of both $\mathbf{S}_{t}^{d}$ and $\mathbf{S}_{t}^{p}$ are drawn from a suitable constellation with average powers of $\sigma^2_d$ and $\sigma^2_p$, respectively, i.e., $\mathbb {E}\left\{ (\mathbf {S}_{t}^{d})(\mathbf {{S}}^{d}_{t})^{H}\right\} =\sigma _{d}^{2} K_1 \mathbf {I}_{M}
\quad \text {and} \quad
\mathrm {Tr}\left ({(\mathbf {S}_{t}^{p})(\mathbf {S}_{t}^{p})^{H} }\right) = \sigma _{p}^{2} MK_{2},$
so that $\sigma _{d}^{2}+\sigma _{p}^{2}=\frac{1}{2}$. In the proposed AP-STS-based system, the data and pilot inputs are affine-precoded over the independent Doppler bins using the semi-orthogonal transmit precoder (TPC) matrices $\mathbf{D}\in\mathbb{C}^{N_{\mathrm{a}}\times K_1}$ and $\mathbf{P}\in\mathbb{C}^{N_{\mathrm{a}}\times K_2}$, which satisfy
\vspace{-2mm}
\begin{align}\label{eq:semi_orth}
\mathbf {P}^{H}\mathbf {P} &= \mathbf {I}_{K_{2}\times K_{2}},~~~~ \mathbf {D}^{H} \mathbf {D}=\mathbf {I}_{K_{1}\times K_{1}}, \nonumber \\
\mathbf {P}^{H}\mathbf {D} &=\mathbf {0}_{K_{2}\times K_{1}},~~~~\mathbf {D}^{H}\mathbf {P}=\mathbf {0}_{K_{1}\times K_{2}}.
\end{align}
Interestingly, TPC matrices such as $\mathbf{P}$ and $\mathbf{D}$ can be readily obtained from an arbitrary unitary matrix $\mathbf {U}\in \mathbb {C}^{N_{\mathrm{a}} \times N_{\mathrm{a}}}$, so that $\mathbf {D}=\mathbf {U}(:,1:K_{1})$ and $\mathbf {P}=\mathbf {U}(:,K_{1}+1:N_{\mathrm{a}})$. Accordingly, the DD-domain AP-STS signal constructed for the $t$th LED is expressed as
\begin{equation}\label{eq:ap_sip_dd_active}
\mathbf{S}_{t,\mathrm{a}}^{\mathrm{dd}}=\mathbf{S}_{t}^{d}\mathbf{D}^{H}+\mathbf{S}_{t}^{p}\mathbf{P}^{H}\in\mathbb{C}^{M\times N_{\mathrm{a}}}.
\end{equation}
To obtain a real-valued time-domain (TD) waveform as required in DCO signaling, Doppler-domain Hermitian symmetry is imposed by construction on the full $N$-length DD grid using $\mathbf{S}_{t,\mathrm{a}}^{\mathrm{dd}}$. In particular, for each delay index $l\in\{0,\ldots,M-1\}$, the entries of $\mathbf{S}_{t}^{\mathrm{dd}}\in\mathbb{C}^{M\times N}$ are defined as \cite{xu2023optical,chen2025optical,zheng2021dco}
\begin{equation}\label{eq:hermitian}
\mathbf{S}_{t}^{\mathrm{dd}}(l,k)=
\begin{cases}
\mathbf{S}_{t,\mathrm{a}}^{\mathrm{dd}}(l,k), & k=1,2,\ldots,\frac{N}{2}-1,\\[2pt]
\big(\mathbf{S}_{t,\mathrm{a}}^{\mathrm{dd}}(l,N-k)\big)^{*}, & k=\frac{N}{2}+1,\ldots,N-1,\\[2pt]
0, & k=0,\frac{N}{2},
\end{cases}
\end{equation}
thereby ensuring Doppler-domain Hermitian symmetry. The DD-domain symbols are subsequently mapped to the TF-domain via the inverse symplectic finite Fourier transform (ISFFT) \cite{xu2023optical}
\begin{align}\label{eq:isfft}
\mathbf{S}_{t}^{\mathrm{tf}}(m,n)
= \frac{1}{\sqrt{NM}}\sum_{l=0}^{M-1}\sum_{k=0}^{N-1}\mathbf{S}_{t}^{\mathrm{dd}}(l,k)
e^{j2\pi\left(\frac{nk}{N}-\frac{ml}{M}\right)},
\end{align}
where $\mathbf{S}_{t}^{\mathrm{tf}}\in\mathbb{C}^{M\times N}$ denotes the TF-domain symbol matrix. Equivalently, \eqref{eq:isfft} admits the matrix representation
\begin{equation}\label{eq:isfft_mat}
\mathbf{S}_{t}^{\mathrm{tf}}=\mathbf{F}_M \mathbf{S}_{t}^{\mathrm{dd}}\mathbf{F}_N^{H},
\end{equation}
with $\mathbf{F}_M$ and $\mathbf{F}_N$ denoting the unitary discrete Fourier transform (DFT) matrices of sizes $M$ and $N$, respectively. Let $p_{\mathrm{tx}}(t)$ denote the transmit pulse of duration $T$. The corresponding TD symbol matrix is then obtained via the Heisenberg transform as \cite{xu2023optical,cai2025power,zheng2021dco}
\begin{align}\label{eq:tx_mat}
\mathbf{X}_{t}=\mathbf{P}_{\mathrm{tx}}\mathbf{F}_M^{H}\mathbf{S}_{t}^{\mathrm{tf}}
=\mathbf{P}_{\mathrm{tx}}\mathbf{S}_{t}^{\mathrm{dd}}\mathbf{F}_N^{H},
\end{align}
where $\mathbf{P}_{\mathrm{tx}}=\mathrm{diag}\{p_{\mathrm{tx}}(\frac{pT}{M})\}_{p=0}^{M-1}\in\mathbb{R}^{M\times M}$. Under the Hermitian symmetry in \eqref{eq:hermitian}, the Doppler-axis inverse DFT (IDFT) yields a real-valued TD waveform. Specifically, upon defining $\tilde{\omega}_{N}=e^{j\frac{2\pi}{N}}$ and letting $\tilde{x}_t[l,n]$ as well as denoting the $(l,n)$th entry of $\mathbf{S}_{t}^{\mathrm{dd}}\mathbf{F}_N^{H}$, we have \cite{xu2023optical,cai2025power,zheng2021dco}
\begin{align}\label{eq:hermitian_property}
\tilde{x}_t[l,n]
&=\frac{1}{\sqrt{N}}\sum_{k=0}^{N-1}\mathbf{S}_{t}^{\mathrm{dd}}(l,k)\tilde{\omega}_{N}^{nk} \nonumber \\
&=\frac{1}{\sqrt{N}}\sum_{k=1}^{\frac{N}{2}-1}\Big(\mathbf{S}_{t}^{\mathrm{dd}}(l,k)\tilde{\omega}_{N}^{nk}+\mathbf{S}_{t}^{\mathrm{dd}}(l,N-k)\tilde{\omega}_{N}^{n(N-k)}\Big) \nonumber \\
&=\frac{1}{\sqrt{N}}\sum_{k=1}^{\frac{N}{2}-1}\Big(\mathbf{S}_{t}^{\mathrm{dd}}(l,k)\tilde{\omega}_{N}^{nk}+\mathbf{S}_{t}^{\mathrm{dd}}(l,k)^{*}\tilde{\omega}_{N}^{-nk}\Big) \nonumber \\
&=\frac{2}{\sqrt{N}}\Re\left\{\sum_{k=1}^{\frac{N}{2}-1}\mathbf{S}_{t}^{\mathrm{dd}}(l,k)\tilde{\omega}_{N}^{nk}\right\},
\end{align}
which implies that $\mathbf{S}_{t}^{\mathrm{dd}}\mathbf{F}_N^{H}\in\mathbb{R}^{M\times N}$ and hence $\mathbf{X}_t\in\mathbb{R}^{M\times N}$ due to the real diagonal weighting $\mathbf{P}_{\mathrm{tx}}$. Subsequently, each column of the TD symbol matrix $\mathbf{X}_t$ is conveyed independently through parallel-to-serial (P/S) conversion, followed by the insertion of a CP of length $L$ to each column. For DCO-OTFS signaling, an appropriate direct current (DC) bias is added to ensure nonnegative optical intensity, while the associated biasing and clipping operations are omitted here for brevity. The next subsection presents the DD-domain VLC channel model under the limited Doppler support (LDS) assumption adopted.

\subsection{DD-Domain VLC Channel Model with Limited Doppler Support (LDS)}
Let $h_{r,t}(\tau,\nu)$ denote the DD-domain VLC channel between the $t$th LED and the $r$th PD, where $1\le r\le N_r$. Owing to the limited number of dominant reflectors in practical VLC environments, the DD-domain channel is modeled as a sparse superposition of discrete paths. Moreover, due to the limited mobility-induced Doppler spread in indoor VLC, the Doppler indices are assumed to lie in an LDS set of size $N_{\nu}\ll N$, where $\mathcal{Q}\triangleq \left\{-\frac{N_{\nu}-1}{2},-\frac{N_{\nu}-1}{2}+1,\ldots,\frac{N_{\nu}-1}{2}\right\}.$
Accordingly, the DD-domain channel is formulated as \cite{hadani2017orthogonal,raviteja2019otfs,sharma2023hyperparameter}
\begin{equation}\label{eq:dd_channel_mimo}
h_{r,t}(\tau,\nu)=\sum_{i=1}^{L_{p}} h_{i,r,t}\,\delta(\tau-\tau_{i})\,\delta(\nu-\nu_{i}),
\end{equation}
where $h_{i,r,t}\in\mathbb{R}_{+}$ denotes the VLC path gain, $\tau_{i}$ and $\nu_{i}$ are the corresponding delay and Doppler shifts, $L_p$ indicates the total number of dominant
multipath components, respectively, and $\delta(\cdot)$ is the Dirac delta function. As commonly adopted in OTFS-based underspread channel modeling \cite{hadani2017orthogonal,raviteja2019otfs}, the multipath delays can be safely assumed to be integer multiples of the delay resolution, i.e., $\tau_i=l_i\Delta\tau$ with $\Delta\tau=\frac{1}{M\Delta f}$, while the Doppler shifts may, in general, be fractional with respect to the Doppler resolution $\Delta\nu=\frac{1}{NT}$. Specifically, the Doppler shift of the $i$th path is expressed as $\nu_i=\frac{k_i}{NT}=k_i\Delta\nu$, where we have $k_i=\mathrm{round}(k_i)+\kappa_i$, and $|\kappa_i|<0.5$. Here, $k_i$ is the (possibly fractional) Doppler index, $\mathrm{round}(k_i)\in\mathbb{Z}$ denotes the nearest integer Doppler bin index, and $\kappa_i$ represents the fractional Doppler offset. Under the LDS assumption, the integer part of the Doppler index is constrained as $\mathrm{round}(k_i)\in\mathcal{Q}$. Furthermore, for a typical underspread channel, the delay and Doppler indices satisfy $l_i\ll M$ and $|k_i|\ll N$ \cite{hadani2017orthogonal,raviteja2019otfs}. Since reflected VLC components experience higher attenuation at longer delays, the power delay profile (PDP) is modeled as an exponentially decaying function across delay bins \cite{saxena2023sparse,liao2023sparse}. Accordingly, the magnitude of the $i$th tap can be expressed as
\begin{equation}\label{eq:exp_decay_mimo}
h_{i,r,t}=
\frac{e^{-l_{i,r,t}\Delta\tau/\tau_{\mathrm{rms},r,t}}}
{\sum_{n=0}^{L_p-1} e^{-n\Delta\tau/\tau_{\mathrm{rms},r,t}}},
\end{equation}
where $l_{i,r,t} \in \{0,1,\ldots,L_p-1\}$ is the delay-bin index of the $i$th path and $\tau_{\mathrm{rms},r,t}$ denotes the channel's root-mean-square (RMS) delay spread \cite{saxena2023sparse,liao2023sparse}. Typically, $\tau_{\mathrm{rms},r,t} \in [0.5\Delta \tau,\,1.5\Delta \tau]$, and its value depends on the average surface reflectivity and room geometry \cite{saxena2023sparse,liao2023sparse}.

\vspace{-4mm}
\subsection{Data-Aided AP-STS MIMO DCO-OTFS Demodulation}
Let $\mathbf {x}_{t,n} \in \mathbb {R}^{M \times 1},~0 \leq n \leq N-1$, denote the $n$th column of the TD symbol matrix $\mathbf{X}_t$ of the $t$th LED, and let $\mathbf {r}_{r,n} \in \mathbb {C}^{M \times 1}$ denote the sample vector received at the $r$th PD. The $p$th sample of $\mathbf {r}_{r,n}$, denoted by $r_{r,n}(p),~0 \leq p \leq M-1$, is expressed as
\begin{equation} \label{C1}
r_{r,n}(p) = \sum _{t=1}^{N_t}\sum _{i=1}^{L_{p}} h_{i,r,t}
e^{j2\pi \frac {{k_{i}}(p-l_{i})}{MN}} x_{t,n}\big ([p-l_{i}]_{M}\big) + w_{r,n}(p),
\end{equation}
where $x_{t,n}(p)$ denotes the $p$th element of the vector $\mathbf{x}_{t,n}$ and $w_{r,n}(p)$ represents the noise samples. Let $\mathbf {r}_{r,n}$ be arranged as $\mathbf {r}_{r,n} = \left [{ r_{r,n}(0), r_{r,n}(1), \cdots, r_{r,n}(M-1) }\right]^{T} \in \mathbb {C}^{M \times 1}$ and $\mathbf{w}_{r,n}$ be arranged as $\mathbf {w}_{r,n} = \left [{ w_{r,n}(0), w_{r,n}(1), \cdots, w_{r,n}(M-1) }\right]^{T}\in \mathbb{C}^{M\times 1}$. Using \eqref{C1}, the received signal vector $\mathbf {r}_{r,n}$ can be formulated as
\begin{align}
\mathbf {r}_{r,n}
=& \sum_{t=1}^{N_t} \sum_{i=1}^{L_{p}} h_{i,r,t}
\left ({\mathbf {\bar \Pi }}\right)^{l_{i}}
\left ({\mathbf {\bar \Delta }_{l_{i},{k_{i}}}}\right) \mathbf {x}_{t,n}
+ \mathbf {w}_{r,n} \nonumber \\
=& \sum _{t=1}^{N_t} \mathbf {\bar H}_{r,t}\mathbf {x}_{t,n} + \mathbf {w}_{r,n},
\end{align}
where the matrix $\mathbf {\bar H}_{r,t} \in \mathbb {C}^{M \times M}$ is defined as
$\mathbf {\bar H}_{r,t} = \sum _{i=1}^{L_{p}} h_{i,r,t} \left ({\mathbf {\bar \Pi }}\right)^{l_{i}} \left ({\mathbf {\bar \Delta }_{{l_{i}},{k_{i}}}}\right)$.
Here, $\mathbf {\bar\Pi}$ denotes a permutation matrix of order $M$ and $\mathbf {\bar \Delta }_{l_{i},{k_{i}}} \in \mathbb {C}^{M \times M}$ is given by \cite{srivastava2021bayesian}
\begin{align}
\bar{\boldsymbol{\Delta}}_{l_i,k_i}=
\begin{cases}
\mathrm{diag}\left\{1,\omega,\ldots,\omega^{M-l_i-1},\omega^{-l_i},\ldots,\omega^{-1}\right\}, & \hspace{-3mm}l_i\neq 0,\\
\mathrm{diag}\left\{1,\omega,\ldots,\omega^{M-1}\right\}, & \hspace{-3mm}l_i=0,
\end{cases}
\end{align}
where $\omega_i = e^{j2\pi \frac {k_i}{MN}}$. Furthermore, upon concatenating the outputs $\mathbf {r}_{r,n},~0 \leq n \leq N-1$, as $\mathbf {R}_r = \left [{ \mathbf {r}_{r,0}, \mathbf {r}_{r,1}, \cdots, \mathbf {r}_{r,N-1}}\right] \in \mathbb {C}^{M \times N}$, we have
\begin{equation}
\mathbf {R}_r = \sum _{t=1}^{N_t} \mathbf {\bar {H}}_{r,t}\mathbf {X}_t + \mathbf {W}_r,
\end{equation}
where $\mathbf {W}_r = \left [{ \mathbf {w}_{r,0}, \mathbf {w}_{r,1}, \cdots, \mathbf {w}_{r,N-1}}\right] \in \mathbb {C}^{M \times N}$ represents the concatenated noise matrix. Subsequently, DCO-OTFS demodulation is applied to the TD sample matrix $\mathbf{R}_r$ as follows.

The DCO-OFDM demodulator first applies a receive pulse-shaping filter $p_{\mathrm{rx}}(t)$ of duration $T$, followed by performing the $M$-point fast Fourier transform (FFT) over each column of the received signal matrix $\mathbf{R}_r$ for obtaining the TF-domain demodulated symbol matrix $\mathbf{Y}^{\mathrm{tf}}_r \in \mathbb{C}^{M\times N}$. These operations are given by
\begin{equation}
\mathbf{Y}^{\mathrm{tf}}_r=\mathbf{F}_M \mathbf{P}_{\mathrm{rx}}\mathbf{R}_r,
\end{equation}
where $\mathbf{P}_{\mathrm{rx}}=\mathrm{diag}\left\{p_{\mathrm{rx}}^{*}\left(\frac{pT}{M}\right)\right\}_{p=0}^{M-1}$. Next, the DD-domain demodulated DCO-OTFS signal $\mathbf{Y}^{\mathrm{dd}}_{r}\in\mathbb{C}^{M\times N}$ is obtained by performing the symplectic FFT (SFFT) of the DCO-OFDM-demodulated signal $\mathbf{Y}^{\mathrm{tf}}_r$, which is expressed as
\begin{equation}
\mathbf{Y}^{\mathrm{dd}}_r=\mathbf{F}_M^{H}\mathbf{Y}^{\mathrm{tf}}_r\mathbf{F}_N
=\mathbf{P}_{\mathrm{rx}}\mathbf{R}_r\mathbf{F}_N.
\end{equation}
Finally, upon substituting $\mathbf{R}_r$, and in turn substituting $\mathbf{X}_t$ into $\mathbf{Y}^{\mathrm{dd}}_r$, the simplified input-output DD-domain relationship of the AP-STS MIMO DCO-OTFS system is expressed as
\begin{equation}\label{eq:8}
\mathbf{Y}^{\mathrm{dd}}_r=\sum _{t=1}^{N_t}\bar{\mathbf{H}}^{\mathrm{dd}}_{r,t}\mathbf{S}^{\mathrm{dd}}_t+\mathbf{W}^{\mathrm{dd}}_r,
\end{equation}
where $\bar{\mathbf{H}}^{\mathrm{dd}}_{r,t}=\mathbf{P}_{\mathrm{rx}}\bar{\mathbf{H}}_{r,t}\mathbf{P}_{\mathrm{tx}}\in\mathbb{C}^{M\times M}$, which can be further reformulated as
\begin{equation}
\bar{\mathbf{H}}^{\mathrm{dd}}_{r,t}
=\sum_{i=1}^{L_p} h_{i,r,t} \mathbf{P}_{\mathrm{rx}}
\left(\bar{\mathbf{\Pi}}\right)^{l_i}
\left(\bar{\mathbf{\Delta}}_{l_i}\right)^{k_i}
\mathbf{P}_{\mathrm{tx}},
\label{eq:9}
\end{equation}
and $\mathbf{W}^{\mathrm{dd}}_r=\mathbf{P}_{\mathrm{rx}}\mathbf{W}_r\mathbf{F}_N$. By stacking the outputs corresponding to all the PDs, the output $\mathbf{Y}^{\mathrm{dd}}\in\mathbb{C}^{MN_r\times N}$ is given by
\begin{equation}
\mathbf{Y}^{\mathrm{dd}}
=\big[ (\mathbf{Y}^{\mathrm{dd}}_1)^{T}\; (\mathbf{Y}^{\mathrm{dd}}_2)^{T}\; \cdots\; (\mathbf{Y}^{\mathrm{dd}}_{N_r})^{T}\big]^{T}.
\end{equation}
Thus, the input-output relationship is given by
\begin{equation}
\mathbf{Y}^{\mathrm{dd}}=\widetilde{\mathbf{H}}^{\mathrm{dd}}\mathbf{S}^{\mathrm{dd}}+\mathbf{W}^{\mathrm{dd}},
\label{eq:48}
\end{equation}
where
$\mathbf{S}^{\mathrm{dd}}
=\big[ (\mathbf{S}^{\mathrm{dd}}_1)^{T}\; (\mathbf{S}^{\mathrm{dd}}_2)^{T}\; \cdots\; (\mathbf{S}^{\mathrm{dd}}_{N_t})^{T}\big]^{T}
\in\mathbb{C}^{MN_t\times N}$,
$\mathbf{W}^{\mathrm{dd}}
=\big[ (\mathbf{W}^{\mathrm{dd}}_1)^{T}\; (\mathbf{W}^{\mathrm{dd}}_2)^{T}\; \cdots\; (\mathbf{W}^{\mathrm{dd}}_{N_r})^{T}\big]^{T}
\in\mathbb{C}^{MN_r\times N}$,
and $\widetilde{\mathbf{H}}^{\mathrm{dd}}\in\mathbb{C}^{MN_r\times MN_t}$ represents the DD-domain MIMO DCO-OTFS channel given as
\begin{align}
\widetilde{\mathbf{H}}^{\mathrm{dd}}
&=\mathrm{blkmtx}\left\{\bar{\mathbf{H}}^{\mathrm{dd}}_{r,t}\right\}_{r=1,t=1}^{N_r,N_t} \notag\\
&=(\mathbf{I}_{N_r}\otimes \mathbf{P}_{\mathrm{rx}})
\Big[\mathrm{blkmtx}\left\{\bar {\mathbf{H}}_{r,t}\right\}_{r=1,t=1}^{N_r,N_t}\Big]
(\mathbf{I}_{N_t}\otimes \mathbf{P}_{\mathrm{tx}}).
\label{eq:49}
\end{align}
Since AP-STS precoding is applied only over the independent Doppler bins, the data-pilot separation is carried out on the corresponding DD-domain output columns. Let \(\mathbf{Y}_{r,\mathrm{a}}^{\mathrm{dd}}\) denote the DD-domain received matrix \(\mathbf{Y}_{r}^{\mathrm{dd}}\) restricted to the independent Doppler bins, and define \(\mathbf{W}_{r,\mathrm{a}}^{\mathrm{dd}}\) similarly. Using the AP-STS construction \(\mathbf{S}_{t,\mathrm{a}}^{\mathrm{dd}}\), the constrained input-output relationship becomes:
\begin{equation}\label{eq:Ydd_a}
\mathbf{Y}_{r,\mathrm{a}}^{\mathrm{dd}}
=\sum_{t=1}^{N_t}\bar{\mathbf{H}}^{\mathrm{dd}}_{r,t}\mathbf{S}_{t,\mathrm{a}}^{\mathrm{dd}}
+\mathbf{W}_{r,\mathrm{a}}^{\mathrm{dd}}.
\end{equation}
Substituting $\mathbf{S}_{t,\mathrm{a}}^{\mathrm{dd}}$ from \eqref{eq:ap_sip_dd_active} into \eqref{eq:Ydd_a} yields
\begin{equation}\label{eq:Ydd_a_expand}
\mathbf{Y}_{r,\mathrm{a}}^{\mathrm{dd}}
=\sum_{t=1}^{N_t}\bar{\mathbf{H}}^{\mathrm{dd}}_{r,t}\left(\mathbf{S}_{t}^{d}\mathbf{D}^{H}+\mathbf{S}_{t}^{p}\mathbf{P}^{H}\right)
+\mathbf{W}_{r,\mathrm{a}}^{\mathrm{dd}}.
\end{equation}
Exploiting the semi-orthogonality of \(\mathbf{D}\) and \(\mathbf{P}\), the decoupled data output is obtained by post-multiplying with \(\mathbf{D}\), namely
\begin{align}
\mathbf{Y}_{r}^{\mathrm{dd},d}
&=\mathbf{Y}_{r,\mathrm{a}}^{\mathrm{dd}}\mathbf{D} \notag\\
&=\sum_{t=1}^{N_t}\bar{\mathbf{H}}^{\mathrm{dd}}_{r,t}
\left(\mathbf{S}_t^{d}\mathbf{D}^{H}\mathbf{D}+\mathbf{S}_t^{p}\mathbf{P}^{H}\mathbf{D}\right)
+\mathbf{W}_{r,\mathrm{a}}^{\mathrm{dd}}\mathbf{D} \notag\\
&=\sum_{t=1}^{N_t}\bar{\mathbf{H}}^{\mathrm{dd}}_{r,t}\mathbf{S}_t^{d}
+\mathbf{W}_{r}^{\mathrm{dd},d}.
\label{eq:51}
\end{align}
where \(\mathbf{W}_{r}^{\mathrm{dd},d}\triangleq \mathbf{W}_{r,\mathrm{a}}^{\mathrm{dd}}\mathbf{D}\). The simplification of \eqref{eq:51} follows upon exploiting the properties in \eqref{eq:semi_orth}. Furthermore, stacking the outputs corresponding to all PDs, the output of the receiver $\widetilde{\mathbf{Y}}^{\mathrm{dd},d}\in\mathbb{C}^{MN_r\times K_1}$ is expressed as
\begin{equation}
\widetilde{\mathbf{Y}}^{\mathrm{dd},d}
=\left[(\mathbf{Y}_{1}^{\mathrm{dd},d})^{T}\; (\mathbf{Y}_{2}^{\mathrm{dd},d})^{T}\; \cdots\; (\mathbf{Y}_{N_r}^{\mathrm{dd},d})^{T}\right]^{T}.
\label{eq:52}
\end{equation}
The system model of decoupled data detection is given by
\begin{equation}
\widetilde{\mathbf{Y}}^{\mathrm{dd},d}
=\widetilde{\mathbf{H}}^{\mathrm{dd}}\widetilde{\mathbf{S}}^{d}
+\widetilde{\mathbf{W}}^{\mathrm{dd},d},
\label{eq:53}
\end{equation}
where
$\widetilde{\mathbf{S}}^{d} =\left[(\mathbf{S}_{1}^{d})^{T} (\mathbf{S}_{2}^{d})^{T} \cdots (\mathbf{S}_{N_t}^{d})^{T}\right]^{T}
\in\mathbb{C}^{MN_t\times K_1}$ and $
\widetilde{\mathbf{W}}^{\mathrm{dd},d}
=\left[(\mathbf{W}_{1}^{\mathrm{dd},d})^{T} (\mathbf{W}_{2}^{\mathrm{dd},d})^{T} \cdots (\mathbf{W}_{N_r}^{\mathrm{dd},d})^{T}\right]^{T}
\in\mathbb{C}^{MN_r\times K_1}$. Once again, the LMMSE-based detector can be formulated as
\begin{align}
\widetilde{\mathbf{S}}^{d}_{\mathrm{LMMSE}}&=\left((\widetilde{\mathbf{H}}^{\mathrm{dd}})^{H}\widetilde{\mathbf{R}}_{{w}_{d}}^{-1}\widetilde{\mathbf{H}}^{\mathrm{dd}}
+\frac{1}{\sigma_d^{2}}\mathbf{I}_{MN_t}\right)^{-1} \nonumber \\
&\times (\widetilde{\mathbf{H}}^{\mathrm{dd}})^{H}\widetilde{\mathbf{R}}_{{w}_{d}}^{-1}
\widetilde{\mathbf{Y}}^{\mathrm{dd},d},
\label{eq:54}
\end{align}
where $\widetilde{\mathbf{R}}_{{w}_{d}}=(\mathbf{I}_{N_r}\otimes \mathbf{R}_{{w}_{d}})\in\mathbb{C}^{MN_r\times MN_r}$ denotes the noise covariance matrix and $\mathbf{R}_{{w}_{d}} = \sigma^2(\mathbf{P}_{\mathrm{rx}}\mathbf{P}_{\mathrm{rx}}^H) \in\mathbb{C}^{M\times M}$ with $\sigma^2$ as the noise variance. The detected symbols are finally demodulated according to the specific transmit constellation using the nearest-neighbor decoding rule.

\section{Sparse DD-domain CIR estimation model for AP-STS MIMO DCO-OTFS VLC Systems}
This section develops a sparse CIR estimation model for the AP-STS MIMO DCO-OTFS VLC system under the LDS assumption. Since the AP-STS precoding is applied only over the independent Doppler bins, the pilot-aided CSI estimation is performed using the DD-domain output restricted to the independent Doppler indices. Let $\mathbf{Y}_{r,\mathrm{a}}^{\mathrm{dd}}\in\mathbb{C}^{M\times N_{\mathrm{a}}}$ denote the DD-domain received matrix at the $r$th PD over the independent Doppler bins, and define $\mathbf{W}_{r,\mathrm{a}}^{\mathrm{dd}}\in\mathbb{C}^{M\times N_{\mathrm{a}}}$ similarly. For CSI estimation, the pilot output $\mathbf{Y}^{\mathrm{dd},p}_r\in\mathbb{C}^{M\times K_{2}}$ is decoupled from the data upon post-multiplying $\mathbf{Y}_{r,\mathrm{a}}^{\mathrm{dd}}$ by the TPC matrix $\mathbf{P}$, i.e., $\mathbf{Y}^{\mathrm{dd},p}_r=\mathbf{Y}_{r,\mathrm{a}}^{\mathrm{dd}}\mathbf{P}$. Using the restricted DD-domain input-output relationship and using \eqref{eq:semi_orth}, we obtain
\begin{align}
\mathbf{Y}_{r}^{\mathrm{dd},p}
=\mathbf{Y}_{r,\mathrm{a}}^{\mathrm{dd}}\mathbf{P}
&=\sum_{t=1}^{N_{t}}
\bar{\mathbf{H}}_{r,t}^{\mathrm{dd}}
\big(
\mathbf{S}_{t}^{d}\mathbf{D}^{H}\mathbf{P}
+\mathbf{S}_{t}^{p}\mathbf{P}^{H}\mathbf{P}
\big)
+\mathbf{W}_{r,\mathrm{a}}^{\mathrm{dd}}
\mathbf{P} \nonumber\\
&=\sum_{t=1}^{N_{t}}\bar{\mathbf{H}}_{r,t}^{\mathrm{dd}}\mathbf{S}_{t}^{p}
+\mathbf{W}_{r}^{\mathrm{dd},p},
\label{eq:55}
\end{align}
where $\mathbf{W}_{r}^{\mathrm{dd},p}\triangleq \mathbf{W}_{r,\mathrm{a}}^{\mathrm{dd}}\mathbf{P}\in\mathbb{C}^{M\times K_2}$. For the DD-domain MIMO DCO-OTFS VLC channel, let the maximum delay spread be denoted by $M_{\tau}$, while the Doppler support is confined to the set $\mathcal{Q}$, where $N_{\nu}=|\mathcal{Q}|\ll N$. For the under-spread channel assumption, the channel parameters satisfy $l_{\max}=\max(l_i)<M_{\tau}\ll M$ and $k_{\max}=\max(k_j)<\frac{N_{\nu}-1}{2}\ll N$. To cater for fractional Doppler effects, a refined Doppler grid of dimension $M_{\tau}\times G_{\nu}$ is considered, with $N_{\nu}=|\mathcal{Q}|\ll G_{\nu}$. In this representation, the $j$th Doppler-grid point, for $0\leq j\leq G_{\nu}-1$, is associated with the Doppler shift $\nu_j = \dfrac{k_j}{NT}$ Hz, where $k_j=-\dfrac{N_{\nu}-1}{2} + \dfrac{j(N_{\nu}-1)}{(G_{\nu}-1)}$. Let $h_{i,j,r,t}$ represent the path gain of the $i$th delay tap and $j$th Doppler tap for the $r$th PD and $t$th LED, given as 
\begin{equation}
h_{r,t}(\tau,\nu)
=
\sum_{i=0}^{M_{\tau}-1}\sum_{j=0}^{G_{\nu}-1}
h_{i,j,r,t}
\delta(\tau-\tau_i)\,\delta(\nu-\nu_j).
\label{eq:56}
\end{equation}
Upon substituting the expression for $\bar{\mathbf{H}}_{r,t}^{\mathrm{dd}}$ into \eqref{eq:55} followed by vectorization, the resultant expression becomes:
\begin{align}
\mathbf{y}_{r}^{\mathrm{dd},p} &=\sum_{t=1}^{N_{t}}
\mathrm{vec}\bigg(\mathbf{P}_{\mathrm{rx}}\bigg(\sum_{i=0}^{M_{\tau}-1}\sum_{j=0}^{G_{\nu}-1}h_{i,j,r,t}(\bar{\boldsymbol{\Pi}})^{i}
\bar{\boldsymbol{\Delta}}_{i,j}\bigg)\mathbf{P}_{\mathrm{tx}}\mathbf{S}_{t}^{p}\bigg) \nonumber \\
& +\mathbf{w}_{r}^{\mathrm{dd},p},
\label{eq:57}
\end{align}
where $\mathbf{y}_{r}^{\mathrm{dd},p}=\mathrm{vec}(\mathbf{Y}_{r,\mathrm{a}}^{\mathrm{dd}}\mathbf{P})\in\mathbb{C}^{MK_{2}\times 1}$,
$\mathbf{w}_{r}^{\mathrm{dd},p}=\mathrm{vec}(\mathbf{W}_{r,\mathrm{a}}^{\mathrm{dd}}\mathbf{P})\in\mathbb{C}^{MK_{2}\times 1}$, and $\mathbf {\bar \Delta }_{i,j} \in \mathbb {C}^{M \times M}$ represents the diagonal matrix, as shown in \cite{srivastava2021bayesian}:
\begin{align}
\mathbf {\bar \Delta }_{i,j}&=\begin{cases} \displaystyle \mathrm {diag}\left \{{1, \bar \omega_j, \cdots, \bar{\omega}^{M-{i}-1}_j, \bar{\omega}^{-{i}}_j, \cdots, \bar{\omega}^{-1}_j }\right\}, \text {if } {i} \ne 0,\\ \displaystyle \mathrm {diag}\left \{{1, \bar \omega_j, \cdots, \bar{\omega}^{M-1}_j }\right \},\text {for } {i}=0, \end{cases} 
\end{align}
where $\bar{\omega}_j = e^{j2\pi \frac {k_j}{MN}}$. Simplifying \eqref{eq:57} yields
\begin{equation}
\mathbf{y}_{r}^{\mathrm{dd},p}
=
\sum_{t=1}^{N_{t}}\sum_{i=0}^{M_{\tau}-1}\sum_{j=0}^{G_{\nu}-1}
\boldsymbol{\omega}_{i,j,t}^{p}h_{i,j,r,t}
+\mathbf{w}_{r}^{\mathrm{dd},p},
\label{eq:58}
\end{equation}
where $\boldsymbol{\omega}_{i,j,t}^{p}
=
\big(\mathbf{I}_{K_{2}}\otimes \mathbf{P}_{\mathrm{rx}}(\bar{\boldsymbol{\Pi}})^{i}(\bar{\boldsymbol{\Delta}}_{i,j})\mathbf{P}_{\mathrm{tx}}\big)\mathbf{s}_{t}^{p}
\in\mathbb{C}^{MK_{2}\times 1}$
and $\mathbf{s}_{t}^{p}=\mathrm{vec}(\mathbf{S}_{t}^{p})\in\mathbb{C}^{MK_{2}\times 1}$.
The above relationship may also be expressed in vectorized form as
\begin{equation}
\mathbf{y}_{r}^{\mathrm{dd},p}
=
\sum_{t=1}^{N_{t}}\boldsymbol{\Omega}_{t}^{p}\mathbf{h}_{r,t}
+\mathbf{w}_{r}^{\mathrm{dd},p},
\label{eq:59}
\end{equation}
where $\mathbf{y}_{r}^{\mathrm{dd},p}$ represents the observation vector for the decoupled pilots at the $r$th PD corresponding to all the LEDs,
$\boldsymbol{\Omega}_{t}^{p}\in\mathbb{C}^{MK_{2}\times M_{\tau}G_{\nu}}$ represents the dictionary matrix for the $t$th LED, which is given by
$\boldsymbol{\Omega}_{t}^{p}=\big[
\boldsymbol{\omega}_{0,0,t}^{p} \cdots \boldsymbol{\omega}_{0,G_{\nu}-1,t}^{p} \cdots
\boldsymbol{\omega}_{M_{\tau}-1,0,t}^{p} \cdots \boldsymbol{\omega}_{M_{\tau}-1,G_{\nu}-1,t}^{p}
\big],$
and the CIR vector
$\mathbf{h}_{r,t}\in\mathbb{R}^{M_{\tau}G_{\nu}\times 1}_{+}$ is given as 
\begin{align}\label{CIR1}
   & \mathbf{h}_{r,t} = \nonumber \\
    &\big[h_{0,0,r,t} \cdots h_{0,G_{\nu}-1,r,t} \cdots h_{M_{\tau}-1,0,r,t} \cdots h_{M_{\tau}-1,G_{\nu}-1,r,t}
\big]^{T}.
\end{align}
The above result can be expressed as
\begin{equation}
\mathbf{y}_{r}^{\mathrm{dd},p}=
\widetilde{\boldsymbol{\Omega}}^{p}\mathbf{h}_{r}
+\mathbf{w}_{r}^{\mathrm{dd},p},
\label{eq:61}
\end{equation}
where the dictionary matrix is given as
$\widetilde{\boldsymbol{\Omega}}^{p}=[\boldsymbol{\Omega}_{1}^{p},\boldsymbol{\Omega}_{2}^{p},\ldots,\boldsymbol{\Omega}_{N_{t}}^{p}]\in\mathbb{C}^{MK_{2}\times M_{\tau}G_{\nu}N_{t}}$,
and the channel coefficient vector for the $r$th PD is given as
$\mathbf{h}_{r}=[\mathbf{h}_{r,1}^{T},\mathbf{h}_{r,2}^{T},\ldots,\mathbf{h}_{r,N_{t}}^{T}]^{T}\in\mathbb{R}^{M_{\tau}G_{\nu}N_{t}\times 1}_{+}$.
Upon concatenating the outputs corresponding to all the PDs using \eqref{eq:61}, the resultant observation matrix
${\mathbf{y}}^{\mathrm{dd},p}=[(\mathbf{y}_{1}^{\mathrm{dd},p})^{T},(\mathbf{y}_{2}^{\mathrm{dd},p})^T,\ldots,(\mathbf{y}_{N_{r}}^{\mathrm{dd},p})^T]^T\in \mathbb{C}^{MK_2N_r\times 1}$
is given by
\begin{equation}
{\mathbf{y}}^{\mathrm{dd},p}
=
{\boldsymbol{\Omega}}^{p}{\mathbf{h}}
+{\mathbf{w}}^{p},
\label{eq:62}
\end{equation}
where the channel coefficient matrix across all the PDs is given by
${\mathbf{h}}=[\mathbf{h}_{1}^T,\mathbf{h}_{2}^T,\ldots,(\mathbf{h}_{N_{r}})^T]^T\in\mathbb{R}^{M_{\tau}G_{\nu}N_{t}N_r\times 1}_{+}$, ${\boldsymbol{\Omega}}^p = I_{N_r} \otimes \widetilde{\boldsymbol{\Omega}}^p \in \mathbb{C}^{MK_2N_r \times M_\tau G_\nu N_t N_r}$
and the corresponding noise matrix is
${\mathbf{w}}^{p}=[(\mathbf{w}_{1}^{\mathrm{dd},p})^T,(\mathbf{w}_{2}^{\mathrm{dd},p})^T,\ldots,(\mathbf{w}_{N_{r}}^{\mathrm{dd},p})^T]
\in\mathbb{C}^{MK_{2}N_{r}\times 1}$.
For the model shown in \eqref{eq:62}, the conventional LMMSE estimate of the CSI, denoted by $\widehat{\mathbf{h}}_{\mathrm{LMMSE}}$, is given as
\begin{equation}
\widehat{\mathbf{h}}_{\mathrm{LMMSE}}
=
\left(({\boldsymbol{\Omega}}^{p})^{H}\mathbf{R}_{w_{p}}^{-1}{\boldsymbol{\Omega}}^{p}+\mathbf{R}_{h}^{-1}\right)^{-1}
({\boldsymbol{\Omega}}^{p})^{H}\mathbf{R}_{w_{p}}^{-1}{\mathbf{y}}^{\mathrm{dd},p},
\label{eq:63}
\end{equation}
where the channel's covariance matrix $\mathbf{R}_{h} = \mathbb{E}[\mathbf{h} {\mathbf{h}}^H]$ is unknown and hence it is set as
$\mathbf{I}_{M_{\tau}G_{\nu}N_{t}N_r}$, while the noise covariance matrix is
$\mathbf{R}_{w_{p}}=\sigma^{2}\big(  \mathbf{I}_{N_r}\otimes\mathbf{I}_{K_{2}}\otimes\mathbf{P}_{\mathrm{rx}}\mathbf{P}_{\mathrm{rx}}^{H}\big)
\in\mathbb{C}^{MK_{2}N_r\times MK_{2}N_r}$. Additionally, another significant drawback of using the conventional
LMMSE-based method for DD-domain CSI estimation is that
it does not exploit the sparse characteristics of the DD-domain channel, which can lead to a significantly improved
performance. To overcome these drawbacks, we conceive a DD-domain
pilot-aided BL (DD-PBL) procedure for enhanced sparse DD-domain CE for a MIMO DCO-OTFS VLC system, as discussed in the next subsection.

\section{DD-domain Pilot-aided BL (DD-PBL) for AP-STS MIMO DCO-OTFS VLC systems}
The proposed DD-PBL technique relies on attributing a parameterized Gaussian prior $f(\mathbf{h};\mathbf{\Upsilon})$ to the sparse multipath VLC CIR vector $\mathbf{h}$, which is unknown. The $k$th component's prior variance of $\mathbf{h}$ is given by the hyperparameter ${\gamma _k}$ and $\mathbf{\Upsilon}= \mathrm {diag}\big \{ \gamma _{k} \big \}_{k=1}^{M_{\tau}G_{\nu}N_tN_r} \in \mathbb{R}^{M_{\tau}G_{\nu}N_tN_r \times M_{\tau}G_{\nu}N_tN_r}$ is the corresponding hyperparameter matrix. When the hyperparameter $\gamma_k$ tends to zero, the equivalent channel component $h_k$ also approaches zero \cite{saxena2023sparse,wipf2004sparse,tipping2001sparse}. Thus, estimating the VLC CIR vector $\mathbf{h}$ is synonymous to calculating the equivalent hyperparameter vector $\boldsymbol{\gamma }=[\gamma _1, \gamma _2, \ldots, \gamma _{M_{\tau}G_{\nu}N_tN_r}]^T$. Moreover, the likelihood function $f(\mathbf{y}^{\mathrm{dd},p}|\mathbf{h})$ is given by 
\begin{align}\label{eqS8}
&f(\mathbf{y}^{\mathrm{dd},p}|\mathbf{h})
=
(\pi)^{-MK_2N_r}\left(\det \left({\mathbf{R}_{w_{p}}}\right)\right)^{-1} \nonumber \\
&\times \exp\left(-\left({\mathbf{y}^{\mathrm{dd},p}-\boldsymbol{\Omega}^{p}\mathbf{h}}\right)^{H}\mathbf{R}_{w_{p}}^{-1}
\left({\mathbf{y}^{\mathrm{dd},p}-\boldsymbol{\Omega}^{p}\mathbf{h}}\right)\right),
\end{align}
where $\mathbf{w}^{p} \sim \mathcal{CN}(\mathbf{0},\mathbf{R}_{w_{p}})$ \cite{saxena2023sparse,wipf2004sparse,tipping2001sparse}. The Bayesian inference is thus provided by
\begin{align}\label{eq:bayes_evidence}
&f(\mathbf{y}^{\mathrm{dd},p}; \mathbf{\Upsilon})
= \int f(\mathbf{y}^{\mathrm{dd},p}|\mathbf{h})f(\mathbf {h};\mathbf{\Upsilon})\,d{\mathbf{h}} =(\pi)^{-MK_2N_r} \nonumber \\
&\times \big(\det(\boldsymbol{\Sigma}_{\mathbf{y}^{\mathrm{dd},p}})\big)^{-1}
\exp\left(-(\mathbf{y}^{\mathrm{dd},p})^{H}\boldsymbol{\Sigma}^{-1}_{\mathbf{y}^{\mathrm{dd},p}}\mathbf{y}^{\mathrm{dd},p}\right),
\end{align}
where the covariance matrix of the output $\mathbf{y}^{\mathrm{dd},p}$ is given as
\begin{equation}\label{eq:sigma_y_ddp}
\boldsymbol{\Sigma }_{\mathbf{y}^{\mathrm{dd},p}}
=\mathbf{R}_{w_{p}} + \boldsymbol{\Omega}^{p} {\mathbf {\Upsilon }} (\boldsymbol{\Omega}^{p})^{H}
\in \mathbb {C}^{MK_2N_r\times MK_2N_r}.
\end{equation}
The hyperparameter vector $\boldsymbol{\gamma}$ is obtained by the maximum-likelihood (ML) estimate, where the optimization objective is determined as follows
\begin{align}\label{eq:98}
\log f(\mathbf{y}^{\mathrm{dd},p};\mathbf{\Upsilon})
&=-MK_2N_r\log\left(\pi\right)-\log\left(\det\left(\boldsymbol{\Sigma}_{\mathbf{y}^{\mathrm{dd},p}}\right)\right)\nonumber \\ 
&-(\mathbf{y}^{\mathrm{dd},p})^{H}\boldsymbol{\Sigma}_{\mathbf{y}^{\mathrm{dd},p}}^{-1}\mathbf{y}^{\mathrm{dd},p}.
\end{align}
The maximization of \eqref{eq:98} is challenging due to its non-concave nature associated with multiple local maxima \cite{wipf2004sparse,tipping2001sparse}. To address this, the EM algorithm is employed iteratively, ensuring an incremental increase in the optimization objective. This method's robust convergence, combined with the specific prior $f(\mathbf {h}; \mathbf \Upsilon)$ in the suggested DD-PBL, enhances the sparse CE performance of MIMO DCO-OTFS VLC systems. The $k$th hyperparameter estimate, in the $m$th iteration, is given by $\widehat{{\gamma }}_k^{(m)}$ and $\mathbf{\widehat{\Upsilon }}^{(m)}=\mathrm{diag}\big \{ \widehat{{\gamma }}_{k}^{(m)} \big \}_{k=1}^{M_{\tau}G_{\nu}N_tN_r}$ represents the corresponding estimated hyperparameter matrix. For the same iteration, the Expectation step (E-step) is employed to compute the log-likelihood ${\mathcal {L}}(\mathbf{\Upsilon}|\widehat{\mathbf{\Upsilon}}^{(m)})$ utilizing the complete dataset $\lbrace \mathbf{y}^{\mathrm{dd},p}, \mathbf{h}\rbrace$, which is given as follows \cite{saxena2023sparse,wipf2004sparse,tipping2001sparse}
\begin{equation} \label{eqS9}
\begin{aligned}
{\mathcal {L}}(\mathbf {\Upsilon }| \widehat{\mathbf {\Upsilon }}^{(m)})
&=\mathbb {E}_{\mathbf {h}|\mathbf{y}^{\mathrm{dd},p}; \widehat{\mathbf {\Upsilon }}^{(m)}}
\Big \lbrace \log f(\mathbf{y}^{\mathrm{dd},p},\mathbf {h}; \mathbf {\Upsilon }) \Big \rbrace \\
&= \mathbb {E}_{\mathbf {h}|\mathbf{y}^{\mathrm{dd},p}; \widehat{\mathbf {\Upsilon }}^{(m)}}
\Big \lbrace \log f(\mathbf {h}; {\mathbf {\Upsilon }}) + \log f(\mathbf{y}^{\mathrm{dd},p}| \mathbf {h}) \Big \rbrace .
\end{aligned}
\end{equation}
It is evident from \eqref{eqS8} that the $f(\mathbf{y}^{\mathrm{dd},p}| \mathbf {h})$ values do not rely on $\mathbf {\Upsilon }$. Consequently, the second term of \eqref{eqS9} can be disregarded, and the Maximization step (M-step) refines the hyperparameter vector $\widehat{\boldsymbol{\gamma}}^{(m)}$ by optimizing the residual objective function in terms of $\boldsymbol{\gamma}$ as \cite{saxena2023sparse,wipf2004sparse,tipping2001sparse}
\begin{align} \label{eqS10}
\widehat{\boldsymbol{\gamma }}^{(m)}
&= \arg \max _{\boldsymbol{\gamma }}
\mathbb {E}_{\mathbf {h}|\mathbf{y}^{\mathrm{dd},p}; \widehat{\mathbf {\Upsilon }}^{(m)}}
\Big \lbrace \log f(\mathbf {h}; {\mathbf {\Upsilon }})\Big \rbrace \nonumber  \\
&= \arg \max _{\boldsymbol{\gamma }}
\sum \limits _{k=1}^{M_{\tau}G_{\nu}N_tN_r}
\Bigg(-\displaystyle \frac{\mathbb {E}_{\mathbf {h}|\mathbf{y}^{\mathrm{dd},p}; \widehat{\mathbf {\Upsilon }}^{(m)}} \lbrace |h(k)|^2  \rbrace}{2\gamma _k} \nonumber \\
& -{\frac{\log (2\pi \gamma _k)}{2}}\Bigg).
\end{align}
In \eqref{eqS10}, the optimization objective can be arranged into $M_{\tau}G_{\nu}N_tN_r $ independent subproblems, each associated with a specific hyperparameter $\gamma_k$. Furthermore, by taking the derivative with respect to $\gamma_k$, followed by equating it to zero yields the estimated hyperparameter $\gamma_k$ as
\begin{equation}\label{eq:gamma_update}
\widehat{\gamma }_k^{(m)}
=\mathbb {E}_{\mathbf {h}|\mathbf{y}^{\mathrm{dd},p}; \widehat{\mathbf {\Upsilon }}^{(m)}}
\Big \lbrace |h(k)|^2 \Big \rbrace
= {\mathbf \Sigma }_{\mathbf h}^{(m)}(k,k) + \big|{\boldsymbol{\mu }}_{\mathbf h}^{(m)}(k)\big|^2.
\end{equation}
The \textit{a posteriori} probability density function of the multipath VLC CIR $\mathbf {h}$ is derived as follows \cite{saxena2023sparse,wipf2004sparse,tipping2001sparse}
\begin{equation}\label{eq:posterior_h}
f(\mathbf {h}|\mathbf{y}^{\mathrm{dd},p}; \widehat{\mathbf {\Upsilon }}^{(m)})
\sim \mathcal {CN}\Big ({\boldsymbol{\mu }}_{\mathbf h}^{(m)}, {\boldsymbol{\Sigma }}_{\mathbf h}^{(m)} \Big),
\end{equation}
so that $
{\boldsymbol{\mu }}_{\mathbf h}^{(m)}
= {\boldsymbol{\Sigma }}_{\mathbf h}^{(m)} (\boldsymbol{\Omega}^{p})^{H} \mathbf{R}_{w_{p}}^{-1} \mathbf{y}^{\mathrm{dd},p}
\in \mathbb {C}^{M_{\tau}G_{\nu}N_tN_r  \times 1}$ is the mean and the covariance matrix is as follows
${\boldsymbol{\Sigma }}_{\mathbf h}^{(m)}
=\bigg((\boldsymbol{\Omega}^{p})^{H} \mathbf{R}_{w_{p}}^{-1} \boldsymbol{\Omega}^{p}
+\left(\widehat{\mathbf {\Upsilon }}^{(m)}\right)^{-1}\bigg)^{-1}
\in \mathbb {C}^{M_{\tau}G_{\nu}N_tN_r  \times M_{\tau}G_{\nu}N_tN_r }.$
Thus, upon convergence, the BL-based method yields the VLC channel's sparse CSI estimate as $\widehat{\mathbf {h}}_{\mathrm{DD-PBL}}={\boldsymbol{\mu }}_{\mathbf h}^{(m)}.$ Subsequently, the estimated ${\widehat {h}}_{i,j,r,t}$ is obtained as $\widehat {h}_{i,j,r,t} = \widehat{\mathbf {h}}_{\mathrm{DD-PBL}}\left[(r-1)N_t M_\tau G_\nu
+(t-1) M_\tau G_\nu+i G_\nu + j + 1\right]$. The estimate $\widehat {\mathbf {H}}_{r,t}^{\mathrm {dd}}$ of the channel matrix corresponding to the $r$th PD and $t$th LED is given by
\begin{equation} \label{nm98}
\widehat {\mathbf {H}}_{r,t}^{\mathrm {dd}}= \mathbf {P}_{\text {rx}}\left[{{ \sum _{i=0}^{M_{\tau }-1} \sum _{j=0}^{G_{\nu }-1} \widehat {h}_{i,j,r,t}\left ({{\bar {\boldsymbol {\Pi }}}}\right)^{i} (\bar {\boldsymbol {\Delta }}_{i})^{j}}}\right] \mathbf {P}_{\text {tx}}. \end{equation}
Subsequently, the estimated CSI $\widehat{\mathbf{H}}_{r,t}^{\mathrm{dd}}$ in \eqref{nm98} is utilized in the LMMSE detector defined in \eqref{eq:54} for data detection. Note that the CSI estimation framework developed above employs only the pilot output $\mathbf{y}^{\mathrm{dd},p}$ for CE. In order to further improve the CSI estimation, one can exploit the estimate of the data symbol matrix $\mathbf{S}^{\mathrm{dd}}$ for iteratively refining the CSI estimate $\widehat{\mathbf{H}}_{r,t}^{\mathrm{dd}}$. Moreover, the resultant uncertainty in the CSI estimates which is characterized by the \textit{a posteriori} covariance matrix $\boldsymbol{\Sigma}^{(m)}$ can be exploited to develop the optimal LMMSE detector. This motivates us to develop a data aided CSI estimation framework for AP-STS MIMO DCO-OTFS VLC systems, which is described next.

\begin{algorithm}[t]
\DontPrintSemicolon
\KwIn{Received pilot vector $\mathbf{y}^{\mathrm{dd},p}$, dictionary matrix $\boldsymbol{\Omega}^{p}$, noise covariance matrix $\mathbf{R}_{w_p}$, stopping parameters $\epsilon$ and $m_{\max}$}
\KwOut{Estimated stacked CSI vector $\widehat{\mathbf{h}}_{\mathrm{DD-PBL}}$}
\textbf{Initialization:} $\widehat{\gamma}_k^{(0)}=1, \forall 1 \le k\le M_{\tau}G_{\nu}N_tN_r,    \widehat{\mathbf{\Upsilon}}^{(0)}=\mathbf{I}_{M_{\tau}G_{\nu}N_tN_r}$
Initialize counter $m=-1$, and $\widehat{\mathbf{\Upsilon}}^{(-1)}=\mathbf{0}$\;
\While{$(\parallel\widehat{\boldsymbol{\gamma }}^{(m+1)} - \widehat{\boldsymbol{\gamma }}^{(m)}\parallel_2 > \epsilon~~ \&\&~~ m < m_{\max})$}
{
$m\leftarrow m+1$

\textbf{E-step:} Compute the \textit{a posteriori} covariance and mean as
$\boldsymbol{\Sigma}_{\mathbf{h}}^{(m)}=\Big(
(\boldsymbol{\Omega}^{p})^{H}\mathbf{R}_{w_p}^{-1}\boldsymbol{\Omega}^{p}+\big(\widehat{\mathbf{\Upsilon}}^{(m-1)}\big)^{-1}\Big)^{-1},$
$\boldsymbol{\mu}_{\mathbf{h}}^{(m)}=
\boldsymbol{\Sigma}_{\mathbf{h}}^{(m)}(\boldsymbol{\Omega}^{p})^{H}\mathbf{R}_{w_p}^{-1}\mathbf{y}^{\mathrm{dd},p}.$

\textbf{M-step:} Update the estimates of the hyperparameters as

\For{$k = 1,2,\ldots,M_{\tau}G_{\nu}N_tN_r$}
{$\widehat{\gamma }_k^{(m)} = {\mathbf \Sigma }_{\mathbf h}^{(m)}(k,k)+|{\boldsymbol{\mu }}_{\mathbf h}^{(m)}(k)|^2$
}\textbf{end}
}\textbf{end}

\textbf{return:~~}{$\widehat{\mathbf{h}}_{\mathrm{DD-PBL}} = {\boldsymbol{\mu }}_{\mathbf h}^{(m)}$}
\caption{DD-PBL-based sparse CSI estimation for MIMO DCO-OTFS VLC systems}
\label{sfblms_algo_dd}
\end{algorithm}

\section{Data aided joint CE and data detection for AP-STS MIMO DCO-OTFS VLC systems}
Commencing with the output $\mathbf{Y}_{r}^{\mathrm{dd},d}$ in \eqref{eq:51} and decoupling the data from the STS, followed by substituting for $\mathbf{H}_{r,t}^{\mathrm{dd}}$ and vectorizing, the resultant expression becomes:
\begin{align}
\mathbf{y}_{r}^{\mathrm{dd},d} &= \sum_{t=1}^{N_{t}} \mathrm{vec}\left(\sum_{i=0}^{M_{\tau}-1}\sum_{j=0}^{G_{\nu}-1}\mathbf{P}_{\mathrm{rx}} h_{i,j,r,t}(\bar{\boldsymbol{\Pi}})^{i}(\bar{\boldsymbol{\Delta}}_{i,j})\mathbf{P}_{\mathrm{tx}}\mathbf{S}_{t}^{d}\right) \nonumber\\
&+\mathbf{w}_{r}^{\mathrm{dd},d},
\label{eq:65}
\end{align}
where $\mathbf{y}_{r}^{\mathrm{dd},d}=\mathrm{vec}(\mathbf{Y}_{r}^{\mathrm{dd},d})\in\mathbb{C}^{MK_{1}\times 1}$ and
$\mathbf{w}_{r}^{\mathrm{dd},d}=\mathrm{vec}(\mathbf{W}_{r}^{\mathrm{dd},d})\in\mathbb{C}^{MK_{1}\times 1}$.
Further simplification of \eqref{eq:65} yields
\begin{equation}
\mathbf{y}_{r}^{\mathrm{dd},d}=
\sum_{t=1}^{N_{t}}\sum_{i=0}^{M_{\tau}-1}\sum_{j=0}^{G_{\nu}-1}\boldsymbol{\omega}_{i,j,t}^{d} h_{i,j,r,t}
+\mathbf{w}_{r}^{\mathrm{dd},d},
\label{eq:66}
\end{equation}
where $\boldsymbol{\omega}_{i,j,t}^{d}=\big(\mathbf{I}_{K_{1}}\otimes \mathbf{P}_{\mathrm{rx}}(\bar{\boldsymbol{\Pi}})^{i}(\bar{\boldsymbol{\Delta}}_{i,j})\mathbf{P}_{\mathrm{tx}}\big)\mathbf{s}_{t}^{d}\in\mathbb{C}^{MK_{1}\times 1}$.
The above relationship can be expressed in the compact form
\begin{equation}
\mathbf{y}_{r}^{\mathrm{dd},d}
=\sum_{t=1}^{N_{t}}\boldsymbol{\Omega}_{t}^{d}\mathbf{h}_{r,t}+\mathbf{w}_{r}^{\mathrm{dd},d},
\label{eq:67}
\end{equation}
where the dictionary matrix $\boldsymbol{\Omega}_{t}^{d}\in\mathbb{C}^{MK_{1}\times M_{\tau}G_{\nu}}$ corresponding to the $t$th LED is given by
\[\boldsymbol{\Omega}_{t}^{d}
=\big[\boldsymbol{\omega}_{0,0,t}^{d}\ \cdots\ \boldsymbol{\omega}_{0,G_{\nu}-1,t}^{d}\ \cdots\
\boldsymbol{\omega}_{M_{\tau}-1,0,t}^{d}\ \cdots\ \boldsymbol{\omega}_{M_{\tau}-1,G_{\nu}-1,t}^{d}\big].\]
The above expression can be further simplified as
\begin{equation}
\mathbf{y}_{r}^{\mathrm{dd},d}=
\widetilde{\boldsymbol{\Omega}}^d\mathbf{h}_{r}
+\mathbf{w}_{r}^{\mathrm{dd},d},
\label{eq:68}
\end{equation}
where $\mathbf{y}_{r}^{\mathrm{dd},d}$ represents the observation vector for the decoupled data at the $r$th PD corresponding to all the LEDs.
Furthermore, the dictionary matrix $\widetilde{\boldsymbol{\Omega}}^{d}\in\mathbb{C}^{MK_{1}\times M_{\tau}G_{\nu}N_{t}}$ is given as
$\widetilde{\boldsymbol{\Omega}}^{d}=[\boldsymbol{\Omega}_{1}^{d},\boldsymbol{\Omega}_{2}^{d},\ldots,\boldsymbol{\Omega}_{N_{t}}^{d}]$,
and $\mathbf{h}_{r}$ for a particular value of $r$ is given by
$\mathbf{h}_{r}=[\mathbf{h}_{r,1}^{T},\mathbf{h}_{r,2}^{T},\ldots,\mathbf{h}_{r,N_{t}}^{T}]^{T}\in \mathbb{R}^{M_{\tau}G_{\nu}N_{t} \times 1}_{+}$.
Upon concatenating the output vectors $\mathbf{y}_{r}^{\mathrm{dd},d}$ corresponding to all the PDs, the resultant observation matrix
$\mathbf{y}^{\mathrm{dd},d}=[(\mathbf{y}_{1}^{\mathrm{dd},d})^T,(\mathbf{y}_{2}^{\mathrm{dd},d})^T,\ldots,(\mathbf{y}_{N_{r}}^{\mathrm{dd},d})^T]^T
\in\mathbb{C}^{MK_{1}N_{r}\times 1}$
can be formulated as
\begin{equation}
\mathbf{y}^{\mathrm{dd},d}
=\boldsymbol{\Omega}^{d}\mathbf{h}+\mathbf{w}^{d},
\label{eq:69}
\end{equation}
where $\mathbf{h}=[\mathbf{h}_{1}^{T},\mathbf{h}_{2}^{T},\ldots,\mathbf{h}_{N_{r}}^{T}]^{T}\in \mathbb{R}^{M_{\tau}G_{\nu}N_{t}N_r \times 1}_{+}$, $\boldsymbol{\Omega}^{d}=\mathbf{I}_{N_r}\otimes \widetilde{\boldsymbol{\Omega}^{d}} \in\mathbb{C}^{MK_{1}N_r\times M_{\tau}G_{\nu}N_tN_r} $, and $\mathbf{w}^{d}=[(\mathbf{w}_{1}^{\mathrm{dd},d})^T,(\mathbf{w}_{2}^{\mathrm{dd},d})^T,\ldots,(\mathbf{w}_{N_{r}}^{\mathrm{dd},d})^T]^T\in\mathbb{C}^{MK_{1}N_{r}\times 1}$. When aiming for data-aided AP-STS-based MIMO-DCO-OTFS CSI estimation, one can stack the outputs from \eqref{eq:62} and \eqref{eq:69} to obtain the joint CSI estimation and detection model
\begin{align}
\underbrace{\begin{bmatrix}
{\mathbf{y}}^{\mathrm{dd},d}\\[0.5mm]
{\mathbf{y}}^{\mathrm{dd},p}
\end{bmatrix}}_{{\mathbf{y}}\in\mathbb{C}^{M N_a N_r\times 1}}
=\underbrace{\begin{bmatrix}
\boldsymbol{\Omega}^{d}\\[0.5mm]
\boldsymbol{\Omega}^{p}
\end{bmatrix}}_{\boldsymbol{\Phi}\in\mathbb{C}^{MN_aN_r\times M_{\tau}G_{\nu}N_{t}N_r}}
\mathbf{h}+\underbrace{\begin{bmatrix}
\mathbf{w}^{d}\\[0.5mm]
\mathbf{w}^{p}
\end{bmatrix}}_{\mathbf{v}\in\mathbb{C}^{MN_aN_r\times 1}}.
\end{align}
Thus, our compact model of DA CSI estimation is given by
\begin{equation}\label{po1}
\mathbf{y}=\boldsymbol{\Phi}\mathbf{h}+\mathbf{v},
\end{equation}
where the covariance matrix of the noise vector $\mathbf{v}$ is $\mathbf {R}_{v}=\mathrm {blkdiag}(\mathbf{R}_{w_{d}}, \mathbf{R}_{w_{p}}) \in \mathbb {C}^{MN_aN_r \times MN_aN_r}$. The DD-DBL technique proposed for joint CSI estimation and data detection proceeds as follows.

For the DD-DBL framework, the parameterized Gaussian prior is assigned to the sparse DD-domain CSI $\mathbf{h}$ given as
\begin{equation}
p(\mathbf{h};\mathbf{\Upsilon})
=\prod_{i=1}^{M_{\tau}G_{\nu}N_{t}N_{r}}\frac{1}{\sqrt{2\pi\lambda_i}}\exp\left(-\frac{|h(i)|^2}{2\lambda_i}\right),
\end{equation}
similar to the DD-PBL framework. The proposed DD-DBL framework jointly and iteratively estimates the hyperparameter matrix $\mathbf{\Upsilon}$ and data symbol matrix $\widetilde{\mathbf {S}}^{d}$ using the EM algorithm. Here, the complete information set is represented as $\{\mathbf{y},\mathbf{h}\}$, where $\mathbf{h}$ is the hidden variable, and $\mathbf{y}$ is the observation variable.
Let $\{\widetilde{\mathbf {S}}^{d},\mathbf{\Upsilon}\}$ represent the unknown parameter set and let $\widehat{\boldsymbol{\eta}}^{(m-1)}=\{{\widehat{\widetilde{\mathbf {S}}}^{d,(m-1)}},\widehat{\mathbf{\Upsilon}}^{(m-1)}\}$, where $\widehat{\widetilde{\mathbf {S}}}^{d,(m-1)}$ and $\widehat{\mathbf{\Upsilon}}^{(m-1)}$ denote the estimates of the data symbol matrix $\widetilde{\mathbf {S}}^{d}$ and the hyperparameter matrix $\mathbf{\Upsilon}$ gleaned from the $(m-1)$st EM iteration. For the current $m$th iteration, the E-step computes the log-likelihood function $\mathcal{L}(\boldsymbol{\eta}| \widehat{\boldsymbol{\eta}}^{(m-1)})$ of the complete data set, which is given by
\begin{align}
&\mathcal{L}(\boldsymbol {\eta}|\widehat {\boldsymbol {\eta}}^{(m-1)})
=\mathbb {E}_{{ {\mathbf {h}}| {\mathbf {y}};\widehat {\boldsymbol {\eta}}^{(m-1)}}}
\left \lbrace{ \log \left [{ f\left ({{\mathbf{y}}, {\mathbf{h}}; \boldsymbol {\eta}}\right)}\right]}\right \rbrace,
\nonumber \\
&=\mathbb {E}\left \lbrace{ \log \left [{ f\left ({{\mathbf{y}}| {\mathbf {h}};\widetilde{\mathbf {S}}^{d}}\right)}\right]}\right \rbrace
+\mathbb {E}\left \lbrace{ \log \left [{f \left ({{\mathbf{h}};\mathbf{\Upsilon}}\right)}\right] }\right \rbrace.
\label{eq:32}
\end{align}
In the M-step, the log-likelihood $\mathcal{L}(\boldsymbol{\eta}|\widehat{\boldsymbol{\eta}}^{(m-1)})$ computed is jointly maximized with respect to the unknown parameter set $\{\widetilde{\mathbf {S}}^{d},\mathbf{\Upsilon}\}$. Note that the first term $\log[f(\mathbf{y}|\mathbf{h};\widetilde{\mathbf {S}}^{d})]$ in \eqref{eq:32} depends only on the data matrix $\widetilde{\mathbf {S}}^{d}$, and the second term $\log[f(\mathbf{h};\mathbf{\Upsilon})]$ depends exclusively on the hyperparameter matrix $\mathbf{\Upsilon}$. Therefore, the joint maximization of $\mathcal{L}(\boldsymbol{\eta}|\widehat{\boldsymbol{\eta}}^{(m-1)})$ with respect to $\widetilde{\mathbf {S}}^{d}$ and $\mathbf{\Upsilon}$ reduces to two independent maximizations of the first and second terms with respect to $\widetilde{\mathbf {S}}^{d}$ and $\mathbf{\Upsilon}$, respectively. The M-step $1$ for the hyperparameter update $\widehat{\gamma}_i^{(m)}$ can be expressed as
\begin{equation}
\widehat {\gamma }^{(m)}_{i}
=\arg \max _{\substack {\gamma_{i}\geq 0}}
\mathbb {E}_{{ \mathbf {h}| {\mathbf{y}};\widehat {\boldsymbol {\eta}}^{(m-1)}}}
\left \lbrace{ \log \left [{ {f}\left ({\mathbf {h};{\mathbf{\Upsilon}}}\right)}\right]}\right \rbrace,
\label{eq:33}
\end{equation}
for $1\leq i\leq M_{\tau}G_{\nu}N_{t}N_{r}$. This can be simplified, similar to Step~$7$ of Algorithm~$1$ as
\begin{equation}
\widehat {\gamma}_{i}^{(m)}
=\boldsymbol {\Sigma }^{(m)}(i,i) + |\widehat {\mathbf {h}}^{(m)}(i)| ^{2},
\label{eq:34}
\end{equation}
where the \textit{a posteriori} mean vector $\widehat{\mathbf{h}}^{(m)}\in\mathbb{C}^{M_{\tau}G_{\nu}N_{t}N_{r}\times 1}$ and the associated covariance matrix $\boldsymbol{\Sigma}^{(m)}\in\mathbb{C}^{M_{\tau}G_{\nu}N_{t}N_{r}\times M_{\tau}G_{\nu}N_{t}N_{r}}$ are determined as
\begin{align}
\widehat {\mathbf {h}}^{(m)}
&=\boldsymbol {\Sigma }^{(m)}\big(\boldsymbol {\Phi }^{(m-1)}\big)^{H} \mathbf {R}_{v}^{-1}\mathbf {y},
\label{eq:35a}\\
\boldsymbol {\Sigma }^{(m)}
&=\Big[\big(\boldsymbol {\Phi }^{(m-1)}\big)^{H} \mathbf {R}_{v}^{-1}\boldsymbol {\Phi }^{(m-1)} + \big(\widehat {\mathbf{\Upsilon}}^{(m-1)}\big)^{-1}\Big]^{-1}.
\label{eq:35b}
\end{align}
Here, the quantity $\boldsymbol{\Phi}^{(m-1)}$ is constructed from the estimate of the data input $\widehat {\widetilde{\mathbf {S}}}^{d,(m-1)}$ at the $(m-1)$st iteration. Subsequently, in M-step $2$, the update $\widehat {\widetilde{\mathbf {S}}}^{d,(m)}$ of the data matrix is given as
\begin{equation}
\widehat {\widetilde{\mathbf {S}}}^{d,(m)}
= \arg \max _{\widetilde{\mathbf {S}}^{d}}
\mathbb {E} \left \{ \log \big[f(\mathbf {y}^{\mathrm{dd,d}}|\mathbf {h}; \widetilde{\mathbf {S}}^{d})\big] \right \},
\label{eq:36}
\end{equation}
which can be further formulated as
\begin{align}
\widehat {\widetilde{\mathbf {S}}}^{d,(m)}
&= \arg \min _{\widetilde{\mathbf {S}}^{d}} \mathbb {E}\big\{ || \mathbf{y}^{\mathrm{dd,d}}-\boldsymbol{\Phi}^d \mathbf {h}||_{2}^{2} \big\} \nonumber \\
&\equiv \arg \min _{\widetilde{\mathbf {S}}^{d}} \mathbb {E}\big\{ || \widetilde{\mathbf {Y}}^{\mathrm{dd,d}} -\mathbf {\widetilde {H}}^{\mathrm{dd},(m)} \widetilde{\mathbf {S}}^{d} ||_{F}^{2} \big\}.
\label{eq:37b}
\end{align}
Upon simplifying the cost function in \eqref{eq:37b}, we obtain
\begin{align}
&\mathbb {E}\Big[\mathrm {Tr} \big\{ (\widetilde{\mathbf {Y}}^{\mathrm{dd,d}} -\mathbf {\widetilde {H}}^{\mathrm{dd},(m)} \widetilde{\mathbf {S}}^{d})^{H}(\widetilde{\mathbf {Y}}^{\mathrm{dd,d}} -\mathbf {\widetilde {H}}^{\mathrm{dd},(m)} \widetilde{\mathbf {S}}^{d}) \big\} \Big]
\nonumber\\
&=\mathrm {Tr} \Big \{ (\widetilde{\mathbf{Y}}^{\mathrm{dd,d}})^{H} (\widetilde{\mathbf {Y}}^{\mathrm{dd,d}})- (\widetilde{\mathbf {Y}}^{\mathrm{dd,d}})^{H} \mathbf {\widehat{H}}^{\mathrm{dd},(m)}\widetilde{\mathbf{S}}^{d} -(\widetilde{\mathbf{S}}^{d})^{H} \times  \nonumber \\
& \big(\mathbf {\widehat{H}}^{\mathrm{dd},(m)}\big)^{H} \widetilde{\mathbf {Y}}^{\mathrm{dd,d}}
+ (\widetilde{\mathbf{S}}^{d})^{H} \mathbb {E}\left[\big(\mathbf {\widetilde{H}}^{\mathrm{dd},(m)}\big)^{H} \big(\mathbf {\widetilde{H}}^{\mathrm{dd},(m)}\big)\right]\widetilde{\mathbf{S}}^{d}\Big \}.
\label{eq:38}
\end{align}
where $\widehat{\mathbf{H}}^{\mathrm{dd, (m)}}$ is the $m$th iteration's CSI estimate, given as
\begin{equation}
\widehat{\mathbf{H}}^{\mathrm{dd, (m)}}
=\mathbb {E} \big [ \widetilde{\mathbf {H}}^{\mathrm{dd},(m)} \big]
= \mathrm{blkmtx}\left\{\widehat{\mathbf{H}}^{\mathrm{dd},(m)}_{r,t}\right\}_{r=1,t=1}^{N_r,N_t},
\label{eq:39}
\end{equation}
where $
\widehat {\mathbf {H}}_{r,t}^{\mathrm {dd}}= \mathbf {P}_{\mathrm {rx}}\left[{{ \sum _{i=0}^{M_{\tau }-1} \sum _{j=0}^{G_{\nu }-1} \widehat {h}_{i,j,r,t}\left ({{\bar {\boldsymbol {\Pi }}}}\right)^{i} (\bar {\boldsymbol {\Delta }}_{i})^{j}}}\right] \mathbf {P}_{\mathrm {tx}}$. The quantity $\widehat {h}_{i,j,r,t}$ is the estimate of $h_{i,j,r,t}$ for the $m$th iteration. The quantity $\mathbb{E}[(\widetilde{\mathbf{H}}^{\mathrm{dd}, (m)})^H(\widetilde{\mathbf{H}}^{\mathrm{dd, (m)}})]$ can be simplified to
\begin{equation}
\mathbb{E}[(\widetilde{\mathbf{H}}^{\mathrm{dd, (m)}})^H(\widetilde{\mathbf{H}}^{\mathrm{dd, (m)}})]
= \big(\widehat {\mathbf {H}}^{\mathrm{dd},(m)}\big)^{H} \widehat {\mathbf {H}}^{\mathrm{dd},(m)} + \boldsymbol {\Xi }^{(m)},
\label{eq:40}
\end{equation}
where $\boldsymbol{\Xi}$ represents the CE uncertainty and it is calculated as follows. Let $\mathbf{h}_{r,t}^{\mathrm{dd}}\in \mathbb{C}^{M^{2}\times 1}$ denote the vectorized equivalent channel corresponding to the $r$th PD and $t$th LED, given as
\begin{align}
\mathbf{h}_{r,t}^{\mathrm{dd}}
= \mathrm{vec}\left(\mathbf{H}_{r,t}^{\mathrm{dd}}\right) &=\mathrm{vec}\left[\sum_{i,j}^{}
h_{i,j,r,t}\mathbf{P}_{\mathrm{rx}}(\bar{\boldsymbol{\Pi}})^{i}\left(\bar{\boldsymbol{\Delta}}_{i}\right)^{j}\mathbf{P}_{\mathrm{tx}}\right] \nonumber\\
&=\sum_{i=0}^{M_{\tau}-1}\sum_{j=0}^{G_{\nu}-1} h_{i,j,r,t}\boldsymbol{\varphi}_{i}^{j},
\label{eq:83}
\end{align}
where $\boldsymbol{\varphi}_{i}^{j}=\mathrm{vec}\left[\mathbf{P}_{\mathrm{rx}}(\bar{\boldsymbol{\Pi}})^{i}\left(\bar{\boldsymbol{\Delta}}_{i}\right)^{j}\mathbf{P}_{\mathrm{tx}}\right]\in\mathbb{C}^{M^{2}\times 1}$.
The above equation can be expressed as $\mathbf{h}_{r,t}^{\mathrm{dd}}=\boldsymbol{\zeta}\mathbf{h}_{r,t}$, where $\boldsymbol{\zeta}\in\mathbb{C}^{M^2\times M_{\tau}G_{\nu}}$ is expressed as
\begin{align} \label{eq:82}
    \boldsymbol{\zeta} = [\boldsymbol{\varphi}_{0}^{0},\cdots,\boldsymbol{\varphi}_{0}^{G_\nu-1},\cdots,\boldsymbol{\varphi}^{0}_{M_\tau-1},\cdots,\boldsymbol{\varphi}^{G_\nu-1}_{M_\tau-1}   ],
\end{align}
and $\mathbf{h}_{r,t}$ is the sparse CIR vector given by \eqref{CIR1}. Upon vectorizing $\widetilde{\mathbf{H}}^{\mathrm{dd}}$, one obtains
\begin{align}
\widetilde{\mathbf{h}}^{\mathrm{dd}}
&=\mathrm{vec}(\widetilde{\mathbf{H}}^{\mathrm{dd}})
=\left[\big(\mathrm{vec}(\mathbf{H}^{\mathrm{dd}}_{1,1})\big)^{T}\ \cdots\ \big(\mathrm{vec}(\mathbf{H}^{\mathrm{dd}}_{N_r,N_t})\big)^{T}\right]^{T} \nonumber\\
&=\left[(\boldsymbol{\zeta}\mathbf{h}_{1,1})^{T},(\boldsymbol{\zeta}\mathbf{h}_{1,2})^{T},\cdots,(\boldsymbol{\zeta}\mathbf{h}_{N_{r},N_{t}})^{T}\right]^{T} \nonumber\\
&=\left(\mathbf{I}_{N_{r}N_{t}}\otimes \boldsymbol{\zeta}\right)
\left[\big(\mathbf{h}_{1,1}\big)^{T},\big(\mathbf{h}_{1,2}\big)^{T},\cdots,\big(\mathbf{h}_{N_{r},N_{t}}\big)^{T}\right]^{T} \nonumber\\
&=\left(\mathbf{I}_{N_{r}N_{t}}\otimes \boldsymbol{\zeta}\right)\widetilde{\mathbf{h}}.
\label{eq:84}
\end{align}
The estimate of $\widetilde{\mathbf{h}}^{\mathrm{dd}}$, denoted by $\widehat{\widetilde{\mathbf{h}}}^{\mathrm{dd}}$, is given as
$\widehat{\widetilde{\mathbf{h}}}^{\mathrm{dd}}=\mathrm{vec}(\widehat{\widetilde{\mathbf{H}}}^{\mathrm{dd}})$.
Let the estimate of $\widetilde{\mathbf{h}}$ be $\widehat{\widetilde{\mathbf{h}}}$.
The relationship between $\widehat{\widetilde{\mathbf{h}}}^{\mathrm{dd}}$ and $\widehat{\widetilde{\mathbf{h}}}$ is given by
$\widehat{\widetilde{\mathbf{h}}}^{\mathrm{dd}}=(\mathbf{I}_{N_{r}N_{t}}\otimes \boldsymbol{\zeta})\widehat{\widetilde{\mathbf{h}}}$,
with the associated error covariance matrix $\boldsymbol{\Sigma}^{(m)}_{h}\in\mathbb{C}^{M^{2}N_rN_t\times M^2N_rN_t}$ determined as
\begin{align}
\boldsymbol{\Sigma}_{h}^{(m)}
=\Big(\mathbf{I}_{N_rN_t}\otimes \boldsymbol{\zeta}\Big)
\boldsymbol{\Sigma}^{(m)}
\left(\mathbf{I}_{N_rN_t}\otimes \boldsymbol{\zeta}^{H}\right),
\label{eq:Sigma_h}
\end{align}
where $\boldsymbol{\Sigma}^{(m)}$ is the error covariance matrix obtained from the E-step of Algorithm \ref{dabl_algo_dd}. It can be shown that for a matrix $\boldsymbol{\Xi}$, the element at $(p,q)$ is given as
\begin{align}
\boldsymbol{\Xi}^{(m)}(p,q) 
&=\mathrm{Tr}\Big[\boldsymbol{\Sigma}_h^{(m)}\big((p-1)MN_r+1:pMN_r,(q-1) \nonumber \\
& \times MN_r+1:qMN_r\big)\Big],
\label{eq:Xi_pq}
\end{align}
where $p,q \in \{ 1, 2, \cdots, MN_t\}.$ Accordingly, the M-step $2$ updates the data matrix by solving the following regularized zero forcing (ZF) problem
\begin{align} 
&\widehat{\widetilde{\mathbf {S}}}^{d,(m)} \nonumber \\ 
&=\arg \min _{\widetilde{\mathbf {S}}^{d}} \left\{{ \Big \Vert\widetilde{\mathbf {Y}}^{\mathrm{dd,d}}-\widehat{\mathbf{H}}^{\mathrm{dd,(m)}}\widetilde{\mathbf {S}}^{d}\Big \Vert_{F}^{2} + \Big \Vert \left ({\boldsymbol {\Xi }^{(m)}}\right)^{\frac {1}{2}} \widetilde{\mathbf{S}}^{d} \Big \Vert_{F}^{2} }\right\} \nonumber \\ 
&=\arg \min _{\widetilde{\mathbf {S}}^{d}}\left\{{ \Bigg \Vert \begin{bmatrix} \widetilde{\mathbf {Y}}^{\mathrm{dd,d}} \\ \mathbf {0} \end{bmatrix} -\begin{bmatrix}\widehat {\mathbf {H}}^{\mathrm{dd},(m)}\\ \left ({\boldsymbol {\Xi }^{(m)}}\right)^{\frac {1}{2}}\\ \end{bmatrix}\widetilde{\mathbf {S}}^{d}\Bigg \Vert _{F}^{2} }\right\}. 
\end{align}
Note that the above detection rule is equivalent to ZF signal detection for the data matrix $\widetilde{\mathbf {S}}^{d}$. Furthermore, the simplified LMMSE-based data detection rule is formulated as
\begin{align}
\widehat{\widetilde{\mathbf{S}}}^{d,(m)}
&=
\Bigg[
\big(\widehat{\mathbf{H}}^{\mathrm{dd,(m)}}\big)^{H}
\widehat{\mathbf{H}}^{\mathrm{dd,(m)}}
+\boldsymbol{\Xi}^{(m)}
+\frac{\sigma^{2}}{\sigma_{d}^{2}}
\big(\widehat{\mathbf{H}}^{\mathrm{dd,(m)}}\big)^{H}\nonumber\\
&\times
\left(\mathbf{I}_{N_r}\otimes\mathbf{P}_{\mathrm{rx}}\mathbf{P}_{\mathrm{rx}}^{H}\right)^{-1}
\widehat{\mathbf{H}}^{\mathrm{dd,(m)}}
\Bigg]^{-1}
\big(\widehat{\mathbf{H}}^{\mathrm{dd,(m)}}\big)^{H}
\widetilde{\mathbf{Y}}^{\mathrm{dd,d}}.
\label{eq:4412}
\end{align}
Under the additional simplification adopted in this work, the detector further reduces to
\begin{align}
\widehat{\widetilde{\mathbf{S}}}^{d,(m)}
&=
\Bigg[
\big(\widehat{\mathbf{H}}^{\mathrm{dd,(m)}}\big)^{H}
\widehat{\mathbf{H}}^{\mathrm{dd,(m)}}
+\boldsymbol{\Xi}^{(m)}
+\frac{\sigma^{2}}{\sigma_{d}^{2}}
\mathbf{I}_{MN_t}
\Bigg]^{-1}  \nonumber\\
& \times \big(\widehat{\mathbf{H}}^{\mathrm{dd,(m)}}\big)^{H}
\widetilde{\mathbf{Y}}^{\mathrm{dd,d}}.
\label{eq:44}
\end{align}
Subsequently, the nearest neighbour detection rule is used to map each element in the matrix above to a symbol in the transmit constellation. The E-step and the M-step, as described above, are executed iteratively until the algorithm converges.
The various steps of the DD-DBL framework are summarized
in Algorithm \ref{dabl_algo_dd}. 
\vspace{-4mm}

\begin{algorithm}[t]
\DontPrintSemicolon
\KwIn{Observation vector $\mathbf{y}$, dictionary matrix $\boldsymbol{\Phi}$, noise covariance matrix $\mathbf{R}_v$, stopping parameters $\epsilon$ and $m_{\max}$}
\KwOut{Estimated CSI vector $\widehat{\mathbf{h}}_{\mathrm{DD-DBL}} = \boldsymbol{\mu}_{\mathbf{h}}^{(m)}$ and $\widehat {\widetilde{\mathbf {S}}}_{\mathrm{DD-DBL}} = \widehat {\widetilde{\mathbf {S}}}^{d,(m)}$}
\textbf{Initialization:} $   \widehat{\mathbf{\Upsilon}}^{(0)}=\widehat{\mathbf{\Upsilon}}^{(m)}_{\mathrm{DD-PBL}}, \widehat{\mathbf{\Upsilon}}^{(-1)}=\mathbf{0}, m=-1$, $(\boldsymbol{\Omega}^{d})^{(-1)}=(\widehat{\boldsymbol{\Omega}}^{d})_{\mathrm{DD-DBL}}$, and $\boldsymbol{\Phi}^{(-1)} = \begin{bmatrix}
(\boldsymbol{\Omega}^{d})^{(-1)}\\[0.5mm]
\boldsymbol{\Omega}^{p}
\end{bmatrix}$

\While{$(\parallel\widehat{\boldsymbol{\gamma }}^{(m+1)} - \widehat{\boldsymbol{\gamma }}^{(m)}\parallel_2 > \epsilon~~ \&\&~~ m < m_{\max})$}
{
$m\leftarrow m+1$

\textbf{E-step:} Compute the \textit{a posteriori} covariance and mean as \\
$\boldsymbol{\Sigma}_{\mathbf{h}}^{(m)}=\Big(
(\boldsymbol{\Phi}^{(m-1)})^{H}\mathbf{R}_v^{-1}\boldsymbol{\Phi}^{(m-1)}+\big(\widehat{\mathbf{\Upsilon}}^{(m-1)}\big)^{-1}\Big)^{-1},$
$\boldsymbol{\mu}_{\mathbf{h}}^{(m)}=
\boldsymbol{\Sigma}_{\mathbf{h}}^{(m)}(\boldsymbol{\Phi}^{(m-1)})^{H}\mathbf{R}_v^{-1}\mathbf{y}.$

\textbf{M-step $1$:} Update the estimates of the hyperparameters as

\For{$k = 1,2,\ldots,M_{\tau}G_{\nu}N_tN_r$}
 {
$\widehat{\gamma }_k^{(m)} = {\mathbf \Sigma }_{\mathbf h}^{(m)}(k,k)+|{\boldsymbol{\mu }}_{\mathbf h}^{(m)}(k)|^2$
}\textbf{end}

\textbf{M-step $2$:}
\vspace{-2mm}
\begin{enumerate}
    \item Update the data estimate $\widehat {\widetilde{\mathbf {S}}}^{d,(m)}$ using \eqref{eq:44}
    \item Demodulate the $\widehat {\widetilde{\mathbf {S}}}^{d,(m)}$ and use it to update $\boldsymbol{\Phi}^{(m)}$
\end{enumerate}

}\textbf{end}

\textbf{return:~~}{$\widehat{\mathbf{h}}_{\mathrm{DD-DBL}} = {\boldsymbol{\mu }}_{\mathbf h}^{(m)}$, $\widehat {\widetilde{\mathbf {S}}}_{\mathrm{DD-DBL}} = \widehat {\widetilde{\mathbf {S}}}^{d,(m)}$}
\caption{DD-DBL-based sparse CSI estimation for MIMO DCO-OTFS VLC systems}
\label{dabl_algo_dd}
\end{algorithm}

\subsection{BCRLB for MIMO DCO-OTFS VLC Systems}
Now the BCRLB is described for characterizing a fundamental lower bound on the achievable mean-square-error (MSE) of any unbiased estimator of the sparse CSI vector $\mathbf{h}$.
Let $\mathbf{T}_B\in \mathbb{C}^{M_{\tau}G_{\nu}N_tN_r\times M_{\tau}G_{\nu}N_tN_r}$ denote the Bayesian Fisher information matrix (BFIM) associated with $\mathbf{h}$. Following standard Bayesian estimation arguments, the BFIM can be expressed as
\vspace{-2mm}
\begin{equation}\label{eq:bfim_decomp}
\mathbf{T}_B=\mathbf{T}_D+\mathbf{T}_P,
\end{equation}
where $\mathbf{T}_D$ is the expected Fisher information matrix (FIM) corresponding to the observation $\mathbf{y}$, while $\mathbf{T}_P$ represents the prior FIM contributed by the channel prior. The data information matrix is defined as
\vspace{-2mm}
\begin{equation}\label{eq:TD_def}
\mathbf{T}_{D}
=-\mathbb{E}_{\mathbf{y},\mathbf{h}}\left\{\frac{\partial^{2}\log f(\mathbf{y}|\mathbf{h})}{\partial \mathbf{h}\partial \mathbf{h}^{H}}\right\}.
\end{equation}
Since $\mathbf{v}\sim \mathcal{CN}(\mathbf{0},\mathbf{R}_v)$, the conditional likelihood is
\begin{align}\label{eq:likelihood_bcrlb}
f(\mathbf{y}|\mathbf{h})
& =(\pi)^{-MN_aN_r}\big(\det(\mathbf{R}_v)\big)^{-1} \nonumber \\
& \times \exp\left(-(\mathbf{y}-\boldsymbol{\Phi}\mathbf{h})^{H}\mathbf{R}_v^{-1}(\mathbf{y}-\boldsymbol{\Phi}\mathbf{h})\right).
\end{align}
Hence, the log-likelihood can be written as
\begin{align}\label{eq:loglik_bcrlb}
\mathcal{L}(\mathbf{y}|\mathbf{h})
=\log f(\mathbf{y}|\mathbf{h}) =C_1-(\mathbf{y}-\boldsymbol{\Phi}\mathbf{h})^{H}\mathbf{R}_v^{-1}(\mathbf{y}-\boldsymbol{\Phi}\mathbf{h}),
\end{align}
where the constant $C_1=-MN_aN_r\log(\pi)-\log\big(\det(\mathbf{R}_v)\big)$ is independent of $\mathbf{h}$. Expanding the quadratic term in \eqref{eq:loglik_bcrlb} yields
\begin{align}\label{eq:loglik_expand_bcrlb}
\mathcal{L}(\mathbf{y}|\mathbf{h})
&=C_1-\mathbf{y}^{H}\mathbf{R}_v^{-1}\mathbf{y}
+\mathbf{h}^{H}\boldsymbol{\Phi}^{H}\mathbf{R}_v^{-1}\mathbf{y}
+\mathbf{y}^{H}\mathbf{R}_v^{-1}\boldsymbol{\Phi}\mathbf{h} \nonumber\\
&-\mathbf{h}^{H}\boldsymbol{\Phi}^{H}\mathbf{R}_v^{-1}\boldsymbol{\Phi}\mathbf{h}.
\end{align}
Upon substituting \eqref{eq:loglik_expand_bcrlb} into \eqref{eq:TD_def}, and observing that only the last term contributes to the second derivative with respect to $\mathbf{h}$ and $\mathbf{h}^H$, we obtain $\mathbf{T}_D=\boldsymbol{\Phi}^{H}\mathbf{R}_v^{-1}\boldsymbol{\Phi}.$ Let us assume a parameterized Gaussian prior for the sparse CSI vector, then the prior FIM is defined as
\begin{align}\label{eq:TP_def}
\mathbf{T}_{P}
=-\mathbb{E}_{\mathbf{h}}\left\{\frac{\partial^{2}\log f(\mathbf{h})}{\partial \mathbf{h}\,\partial \mathbf{h}^{H}}\right\}.
\end{align}
The log-prior can be expressed as
\begin{align}\label{eq:logprior_bcrlb}
\mathcal{L}(\mathbf{h};\mathbf{\Upsilon})
&=\log f(\mathbf{h})
=C_2-\mathbf{h}^{H}\mathbf{\Upsilon}^{-1}\mathbf{h},
\end{align}
where $C_2=-M_{\tau}G_{\nu}N_tN_r\log(\pi)-\log\big(\det(\mathbf{\Upsilon})\big)$
is independent of $\mathbf{h}$. Substituting \eqref{eq:logprior_bcrlb} into \eqref{eq:TP_def} yields $\mathbf{T}_P=\mathbf{\Upsilon}^{-1}.$ Thus, the BFIM is given as
\begin{align}\label{eq:TB_result}
\mathbf{T}_B=\mathbf{\Upsilon}^{-1}+\boldsymbol{\Phi}^{H}\mathbf{R}_v^{-1}\boldsymbol{\Phi}.
\end{align}
Accordingly, the BCRLB on the MSE of any unbiased estimator $\widehat{\mathbf{h}}$ becomes:
\begin{align}\label{eq:BCRLB_h}
\mathrm{MSE}(\widehat{\mathbf{h}})
=\mathbb{E}\{||\widehat{\mathbf{h}}-\mathbf{h}||_2^2\}
&\geq \mathrm{Tr}\left(\mathbf{T}_B^{-1}\right) \nonumber\\
&=\mathrm{Tr}\left(\left(\mathbf{\Upsilon}^{-1}+\boldsymbol{\Phi}^{H}\mathbf{R}_v^{-1}\boldsymbol{\Phi}\right)^{-1}\right).
\end{align}
Furthermore, using the DD-domain relationship, the BCRLB for the MSE of the DD-domain CE $\widehat{\widetilde{\mathbf{h}}}^{\mathrm{dd}}$ can be expressed as
\begin{align}\label{eq:BCRLB_hdd}
\mathrm{MSE}\left(\widehat{\widetilde{\mathbf{h}}}^{\mathrm{dd}}\right)
&\geq
\mathrm{Tr}\left\{\left(\mathbf{I}_{N_rN_t}\otimes \boldsymbol{\zeta}\right)\mathbf{T}_B^{-1}
\left(\mathbf{I}_{N_rN_t}\otimes \boldsymbol{\zeta}\right)^{H}\right\}.
\end{align}

\begin{figure*}[ht]
\centering
\captionsetup[subfigure]{justification=centering}
\subfloat[]{\label{1}\includegraphics[width=75mm,height=56mm]{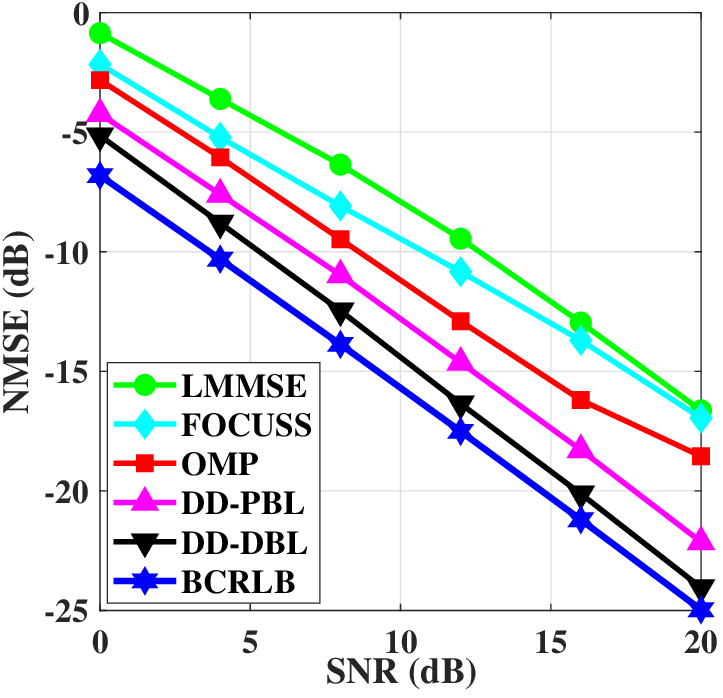}} \hspace{10mm}
\subfloat[]{\label{2}\includegraphics[width=75mm,height=56mm]{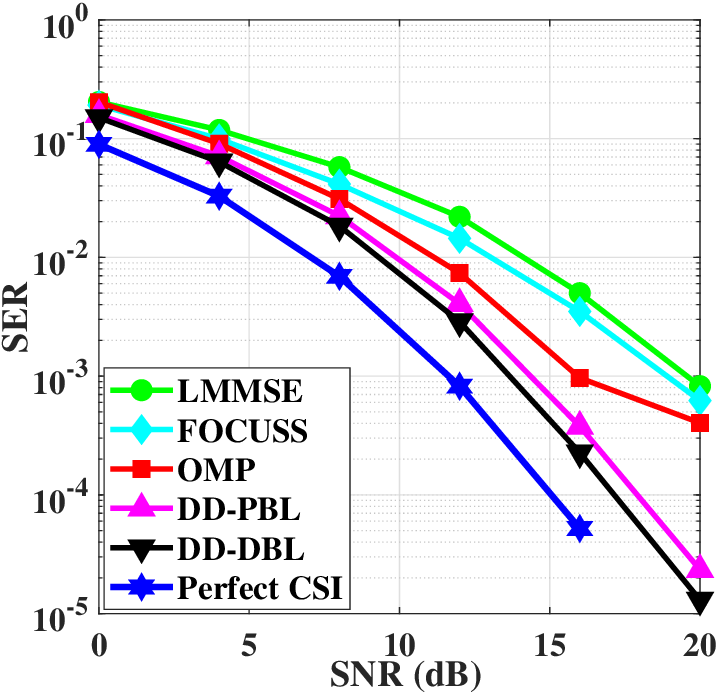}}
\caption{AP-STS MIMO DCO-OTFS VLC system, demonstrates (a) NMSE versus SNR performance;  (b) SER versus SNR performance.}
\label{a}
\vspace{-1mm}
\end{figure*}

\begin{figure*}[!htb]
\centering
\captionsetup[subfigure]{justification=centering}
\subfloat[]{\label{3}\includegraphics[width=75mm,height=56mm]{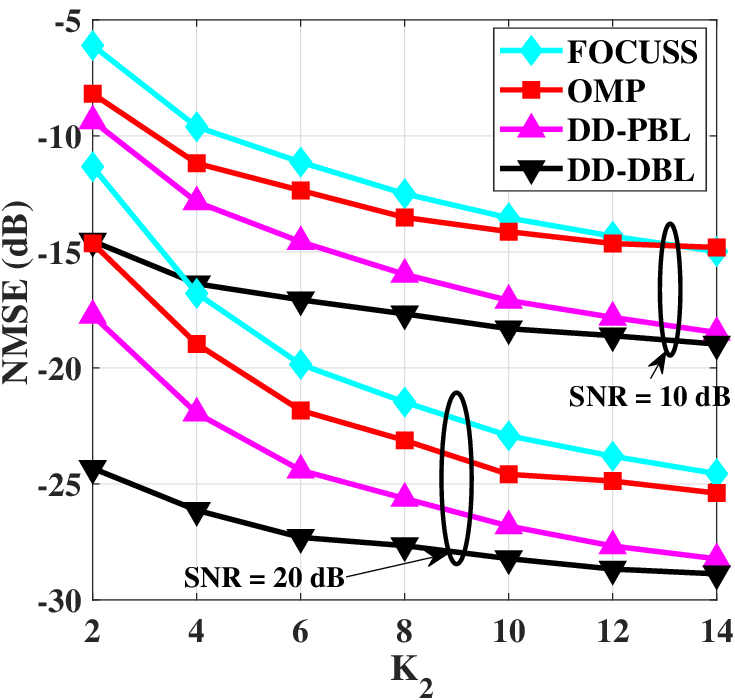}}\hspace{10mm}
\subfloat[]{\label{4}\includegraphics[width=75mm,height=56mm]{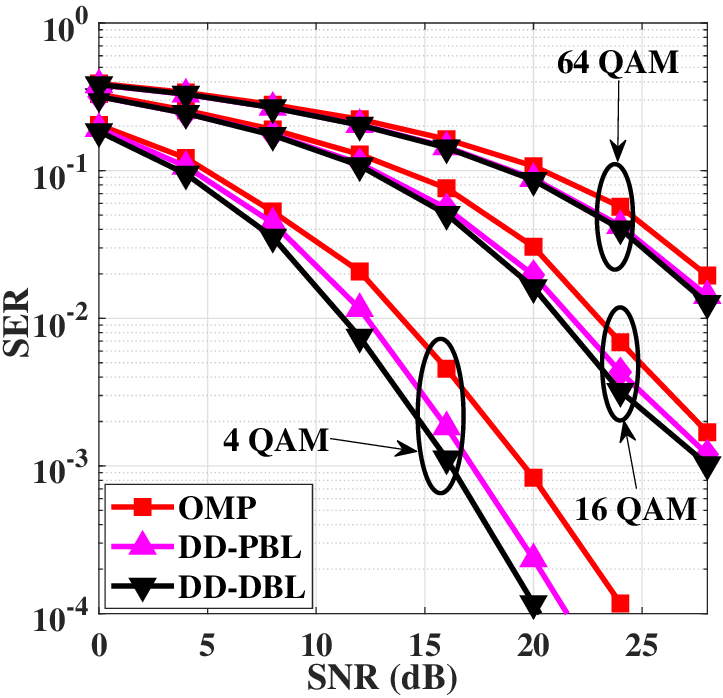}}
\caption{AP-STS MIMO DCO-OTFS VLC system, demonstrates (a) NMSE versus pilot length ($K_2$) performance at SNR $= 10$ and $20$ dB; (b) SER versus SNR performance with higher-order modulation.}
\label{b}
\vspace{-2mm}
\end{figure*}
\vspace{-4mm}
\section{Simulation Results}
This section presents a comparative performance analysis of the proposed DD-PBL and DD-DBL-based method against existing CIR estimation techniques, namely FOCUSS, OMP, and LMMSE \cite{saxena2023sparse}, for CSI recovery in AP-STS MIMO DCO-OTFS VLC systems. The evaluation considers the SER, NMSE, and pilot length as key performance indicators. The SER quantifies detection reliability at the receiver using the estimated CIR. In the simulation setup, $\mathbf{R}_h = \mathbf{I}_{M_\tau N_\nu N_t N_r}$, while the SNR (in dB) is defined as $\text{SNR (dB)} = 10\log_{10}\left(\frac{1}{\sigma^2}\right).$ The DD-PBL and DD-DBL algorithms employ the convergence parameters of $\epsilon = 10^{-6}$ and $m_{\max} = 50$, whereas the OMP threshold is fixed at $\xi = 0.1$. The FOCUSS configuration adopts a noise regularization factor $\sigma^2$, utilizes the $l_p$-norm with $p=0.8$, and applies a termination criterion of $10^{-5}$ with a maximum of $800$ iterations. The additional simulation parameters are: $\Delta f = 240$ kHz, $M= 64$, $N= 64$, $K_2 = 4$, $L= 16$, $L_p = 5$, $M_\tau = 16$, $N_\nu = 15$, $N_r =2$, $N_t =2$, and $B_{\text{DC}} = 7$ dB. The modulation scheme is binary phase shift keying (BPSK), and pulse shape is rectangular \cite{saxena2023sparse,liao2023sparse,xu2023optical}.

\begin{figure*}[ht]
\centering
\captionsetup[subfigure]{justification=centering}
\subfloat[]{\label{5}\includegraphics[width=75mm,height=56mm]{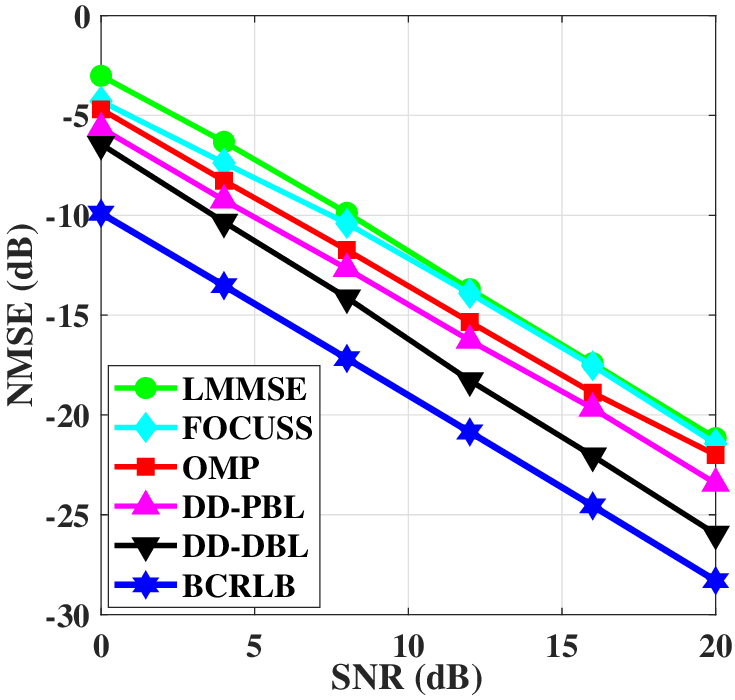}} \hspace{10mm}
\subfloat[]{\label{6}\includegraphics[width=75mm,height=56mm]{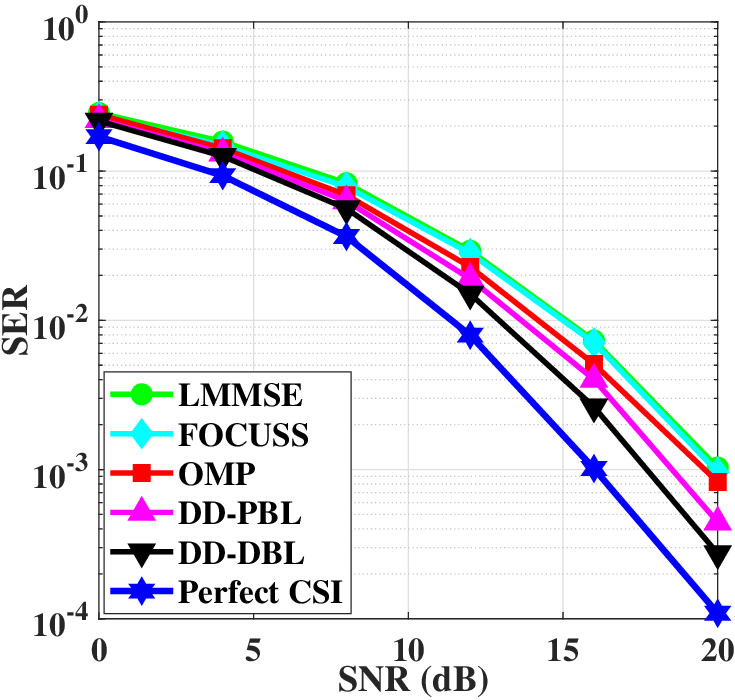}}
\caption{AP-STS MIMO DCO-OTFS VLC system, demonstrates (a) NMSE versus SNR performance;  (b) SER versus SNR performance.}
\label{c}
\vspace{-1mm}
\end{figure*}

\begin{figure*}[ht]
\centering
\captionsetup[subfigure]{justification=centering}
\subfloat[]{\label{7}\includegraphics[width=75mm,height=56mm]{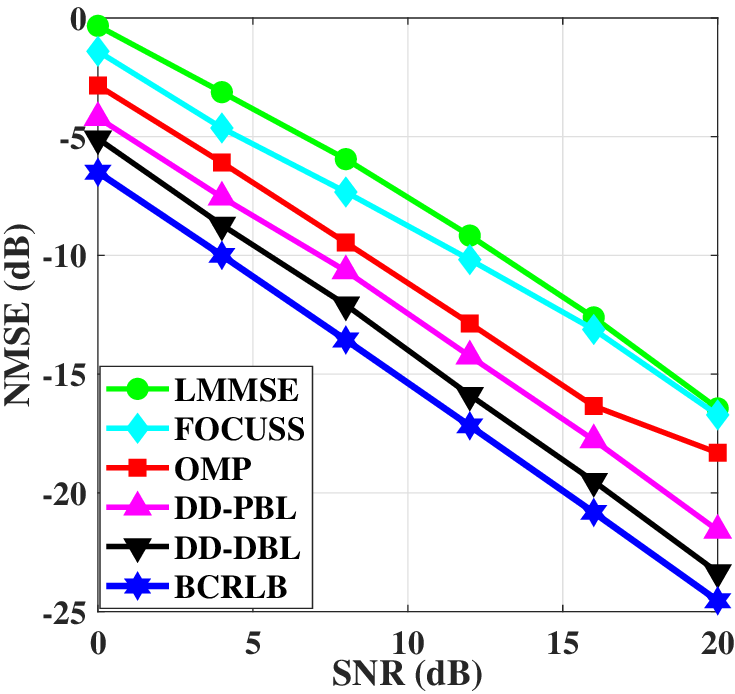}} \hspace{10mm}
\subfloat[]{\label{8}\includegraphics[width=75mm,height=56mm]{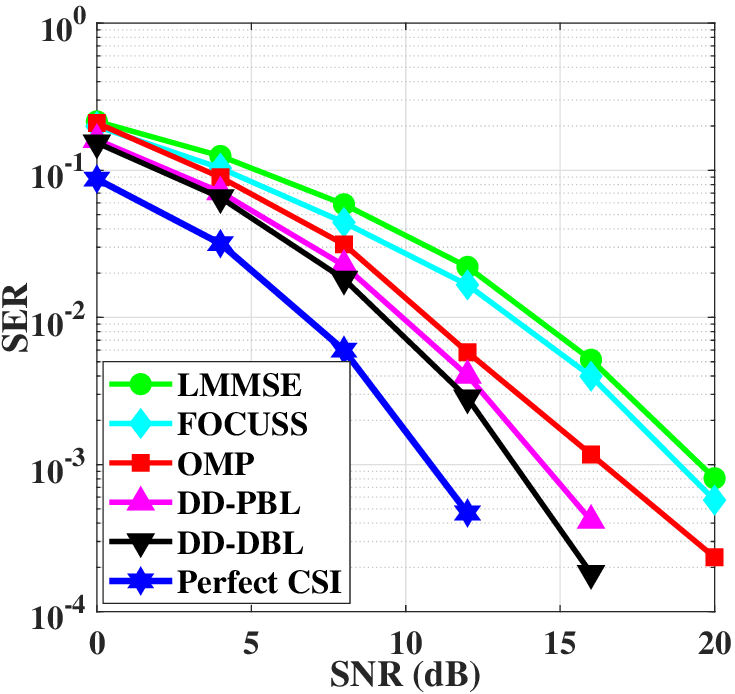}}
\caption{AP-STS MIMO DCO-OTFS VLC system with fractional Doppler, demonstrates (a) NMSE versus SNR performance; (b) SER versus SNR performance.}
\label{d}
\vspace{-2mm}
\end{figure*}
Fig. \ref{a}\subref{1} compares the NMSE performance of the proposed DD-PBL and DD-DBL estimators to the benchmark schemes. The NMSE is defined as NMSE$~= \frac{||\widehat{\mathbf{h}}-\mathbf{h}||_2^2}{||\mathbf{h}||_2^2}$. As observed in Fig. \ref{a}\subref{1}, the BL-based DD-PBL and DD-DBL achieve lower NMSE than OMP, FOCUSS, and LMMSE for the AP-STS MIMO DCO-OTFS VLC systems. The relatively weak performance of OMP is primarily attributable to its reliance on an empirically selected stopping criterion. By contrast, FOCUSS suffers from convergence limitations and pronounced sensitivity to the regularization parameter, which degrades its robustness \cite{saxena2023sparse}. Since the conventional LMMSE estimator does not exploit the DD-domain CSI sparsity, it exhibits the poorest NMSE behavior. Overall, the non-Bayesian sparse recovery techniques, namely OMP and FOCUSS, are less reliable than the BL-based estimators due to the aforementioned shortcomings. Among the schemes considered, DD-DBL achieves the most favorable performance. This gain stems from its ability to incorporate data estimates obtained using the modified LMMSE rule in \eqref{eq:44}, in addition to operating with limited pilot overhead. Notably, DD-DBL approaches the BCRLB at high SNR values, despite dispensing with any prior knowledge of the channel covariance matrix, which is typically essential for conventional LMMSE processing. Moreover, it does not presume knowledge of the sparse support. These properties highlight the practical appeal of DD-DBL for DCO-OTFS implementations, where reliable prior information is often unavailable. The joint CSI estimation and detection capability further strengthens performance by exploiting the abundant data symbols alongside the comparatively small number of pilots, leading to a behavior that closely tracks the BCRLB in the high-SNR regime.

Fig. \ref{a}\subref{2} further characterizes the SER results obtained when detecting the superimposed data symbols using the CSI delivered by each estimator considered. In particular, the SER is computed for the data symbols embedded onto the pilots by utilizing the CSI produced by the previously discussed estimation methods. The corresponding performance is also compared to that of an idealized receiver operating with perfect CSI. As expected, the BL-based approaches, namely DD-PBL and DD-DBL, attain lower SER than the non-BL baselines OMP and FOCUSS, which is consistent with the NMSE trends observed in Fig. \ref{a}\subref{1}. Additionally, DD-DBL again provides the most favorable SER performance, with its curve closely approaching that of the perfect-CSI benchmark. These results confirm the effectiveness of the proposed DD-DBL strategy in producing highly accurate CSI estimates for reliable data detection.

Fig. \ref{b}\subref{3} depicts the NMSE as a function of the pilot length $K_2$ for the proposed sparse CSI estimation schemes in an AP-STS MIMO DCO-OTFS VLC system at $\mathrm{SNR}=10$ dB and $20$ dB. The results show a monotonic reduction in NMSE as $K_2$ increases, which is consistent with the role of training in improving estimation quality. Specifically, enlarging $K_2$ increases the number of available observations, thereby lowering the estimation error \cite{kay1993fundamentals}. Among the techniques considered, DD-DBL again attains the lowest NMSE, highlighting the benefit of additionally exploiting the data symbols. In the proposed design, $MK_2N_t$ pilot symbols are embedded within a block of $MN_aN_t$ symbols, which yields a pilot overhead of $\rho=\frac{K_2}{N_a}$. By contrast, the embedded-pilot (EP) approach requires a substantially higher overhead, given approximately by $\rho_{\mathrm{EP}}\approx \frac{(N_t M_\tau + M_\tau + N_t)(2N_{\nu}+1)}{MN_aN_t}$ \cite{raviteja2019embedded}. For the parameters considered and $K_2 = 4$, this results in $\rho_{\mathrm{EP}}=0.42$ versus $\rho=0.13$, demonstrating that the proposed scheme maintains high bandwidth efficiency.

Fig. \ref{b}\subref{4} compares the SER of DD-DBL, DD-PBL, and OMP for $4$, $16$, and $64$ QAM signaling. In all cases, DD-DBL achieves the lowest SER, indicating more reliable detection under the CSI it produces. These results show that the proposed DD-DBL remains effective across modulation orders and consistently improves over both DD-PBL and OMP in the SNR range considered.

To highlight the merits of the proposed framework, the DD-DBL and DD-PBL schemes advocated are assessed under multiple simulation settings with $\Delta f=480$ kHz, $M=32$, $N=64$, $K_2=6$, $L= 8$, $L_p = 4$, $M_\tau = 8$, $N_\nu = 7$, 
while all other parameters are kept unchanged. Fig. \ref{c}\subref{5} compares the NMSE achieved by the proposed estimators to that of representative benchmark methods for the AP-STS MIMO DCO-OTFS VLC system, where DD-DBL consistently delivers the best performance. Fig. \ref{c}\subref{6} reports the corresponding SER results and shows that DD-DBL attains substantially lower SER than DD-PBL, OMP, and FOCUSS, which is attributed to its improved CSI quality and its ability to approach the performance of a receiver operating with perfect CSI.

Figs. \ref{d}\subref{7} and \ref{d}\subref{8} illustrate the NMSE and SER attained by the considered algorithms for AP-STS MIMO DCO-OTFS VLC links with $G_\nu = 32$, which captures fractional Doppler effects. The results show that DD-DBL consistently provides superior performance relative to the benchmark estimators, reflecting its improved CSI quality and the associated detection reliability. Moreover, increasing the number of Doppler bins $G_\nu$ enhances the DD-grid resolution, which further improves both the estimation and detection performance.
\vspace{-3mm}
\section{Conclusions}
This work presented an orthogonal AP-STS framework for CP-assisted MIMO DCO-OTFS VLC links for transmission over doubly selective channels. By jointly affine-precoding and superimposing pilot and data matrices in the DD domain, a unified end-to-end DD-domain input-output model was derived, while orthogonal precoders at each PD enabled reliable pilot/data separation with suppressed mutual interference. Based on this model, an EM-driven DD-PBL algorithm was developed for CSI estimation, followed by a DD-DBL procedure that iteratively refines CSI and detects data by leveraging detected symbols as virtual pilots. The resultant LMMSE detector explicitly incorporates CSI uncertainty, and the BCRLB was derived for the setting considered. Numerical results demonstrate improved NMSE, reduced pilot overhead, and mitigated SER compared to the benchmark schemes.
\vspace{-4mm}
\bibliographystyle{IEEEtran}
\bibliography{citation}

\end{document}